\documentclass[iop]{emulateapj}
\usepackage{apjfonts}
\usepackage{natbib}

\usepackage{epsfig}
\usepackage{amssymb}
\usepackage{natbib}
\usepackage{aas_macros}
\usepackage{amsmath}

\DeclareGraphicsExtensions{.eps}

\newcommand{\gamm}{$\gamma'$}
\newcommand{\gamme}{\gamma'}

\newcommand{\app}{Appendix~}
\newcommand{\figu}{Figure~}
\newcommand{\figus}{Figures~}
\newcommand{\eq}{Equation~}
\newcommand{\sect}{Section~}

\def\mctwoe{M_{\rm 200c}}

\newcommand{\devac}{de Vaucouleurs}

\newcommand{\re}{$R_{\rm e}$}
\newcommand{\ree}{R_{\rm e}}

\newcommand{\mstart}{$M_{\rm star}$}

\newcommand{\mhaloe}{M_{\rm halo}}

\newcommand{\msune}{M_{\odot}}

\newcommand{\mstare}{M_{\rm star}}

\newcommand{\kpce}{{\rm kpc}}

\begin{document}

\voffset=-1.00in

\def\sarc{$^{\prime\prime}\!\!.$}
\def\arcsec{$^{\prime\prime}$}
\def\arcmin{$^{\prime}$}
\def\degr{$^{\circ}$}
\def\seco{$^{\rm s}\!\!.$}
\def\ls{\lower 2pt \hbox{$\;\scriptscriptstyle \buildrel<\over\sim\;$}}
\def\gs{\lower 2pt \hbox{$\;\scriptscriptstyle \buildrel>\over\sim\;$}}

\title{Revisiting the Bulge-Halo Conspiracy I: Dependence on galaxy properties and halo mass}

\author{Francesco Shankar\altaffilmark{1},
Alessandro Sonnenfeld\altaffilmark{2,3}, Gary A. Mamon\altaffilmark{4}, Kyu-Hyun Chae\altaffilmark{5}, Raphael Gavazzi\altaffilmark{4}, Tommaso Treu\altaffilmark{2,3}, Benedikt Diemer\altaffilmark{6}, Carlo Nipoti\altaffilmark{7}, Stewart Buchan\altaffilmark{1}, Mariangela Bernardi\altaffilmark{8}, Ravi Sheth\altaffilmark{8},
Marc Huertas-Company\altaffilmark{9,10}}
\altaffiltext{1}{Department of Physics and Astronomy, University of Southampton, Highfield, SO17 1BJ, UK; F.Shankar@soton.ac.uk}
\altaffiltext{2}{Physics Department, University of California, Santa Barbara, CA 93106, USA}
\altaffiltext{3}{Physics and Astronomy Department University of California Los Angeles CA 90095-1547}
\altaffiltext{4}{Institut d'Astrophysique de Paris (UMR 7095: CNRS \& UPMC, Sorbonne Universit\'{e}s), 98bis Bd Arago, F-75014 Paris, France}
\altaffiltext{5}{Department of Physics and Astronomy, Sejong University, 209 Neungdong-ro Gwangjin-Gu, Seoul 05006, Republic of Korea}
\altaffiltext{6}{Institute for Theory and Computation, Harvard-Smithsonian Center for Astrophysics, 60 Garden St., Cambridge MA 02138}
\altaffiltext{7}{Department of Physics and Astronomy, Bologna University, viale Berti-Pichat 6/2, I-40127 Bologna, Italy}
\altaffiltext{8}{Department of Physics and Astronomy, University of Pennsylvania, 209 South 33rd St, Philadelphia, PA 19104}
\altaffiltext{9}{GEPI, Observatoire de Paris, CNRS, Universit\'e Paris Diderot, 61, Avenue de l'Observatoire 75014, Paris  France}
\altaffiltext{10}{Universit\'{e} Paris Denis Diderot, 75205 Paris Cedex 13, France}

\begin{abstract}
We carry out a systematic investigation of the total mass density profile of massive ($\log \mstare/\msune \gtrsim 11.3$) early-type galaxies and its dependence on galactic properties and host halo mass with the aid of a variety of lensing/dynamical data and large mock galaxy catalogs.
The latter are produced via semi-empirical models that, by design, are based on just a few basic input assumptions. Galaxies, with measured stellar masses, effective radii and S\'{e}rsic indices, are assigned, via abundance matching relations, host dark matter halos characterized by a typical $\Lambda$CDM profile. Our main results are as follows: (i) In line with observational evidence, our semi-empirical models naturally predict that the total, mass-weighted density slope at the effective radius \gamm\ is not universal, steepening for more compact and/or massive galaxies, but flattening with increasing host halo mass. (ii) Models characterized by a Salpeter or variable initial mass function and uncontracted dark matter profiles are in good agreement with the data, while a Chabrier initial mass function and/or adiabatic contractions/expansions of the dark matter halos are highly disfavored. (iii) Currently available data on the mass density profiles of very massive galaxies ($\log \mstare/\msune \gtrsim 12$), with $\mhaloe \gtrsim 3\times 10^{14}\, \msune$, favor instead models with a stellar profile flatter than a S\'{e}rsic one in the very inner regions ($r\lesssim 3-5$ kpc), and a cored NFW or Einasto dark matter profile with median halo concentration a factor of $\sim 2$ or $\lesssim 1.3$, respectively, higher than those typically predicted by N-body numerical simulations.
\end{abstract}


\section{Introduction}
\label{sec|intro}

Early-type galaxies constitute a family of objects of remarkable regularity,
captured by a series of tight scaling relations such as the fundamental plane \citep[e.g.,][]{Dressler87,Djo87}, the color-magnitude relation \citep[e.g.][]{Bower92,Mei12} and the relation between the mass of the central black hole and the global properties of the host galaxy \citep[e.g.,][]{Gebhardt00,Ferrarese00,ShankarReview,Shankar16BH}.
This regularity is reflected in their mass structure: the total mass density profile of massive\footnote{This is not necessarily the case if one considers a broader mass range \citep[e.g.,][]{Shu15}.} galaxies is well approximated, around the scale of the half-light radius, by a power-law $\rho(r)\propto r^{-\gamma}$ with slope close to isothermal ($\gamma \approx 2$) and little scatter across the population \citep[e.g.,][]{Koop06,Gavazzi07,Koop09,Barnabe11}.
This compelling observation takes the name of {\em bulge-halo conspiracy}: stars and dark matter have separately density profiles significantly different from an isothermal profile, yet they conspire to produce total density profile close to isothermal \citep[e.g.,][]{TreuKoopmans04}.

Recent studies based on strong lensing and stellar dynamics have shown how the density profile of massive early-type galaxies is not exactly universal: the density slope \gamm\ correlates with projected stellar mass density, being steeper for more compact objects, and anti-correlates with redshift, being shallower for higher redshift object with respect to local systems of the same mass and size \citep[e.g.,][]{Auger10,Ruff11,Bolton12,Sonne13,Dye14,Tortora14a,Tortora14b}.
The slope of the density profile appears to also correlate with halo mass: cluster Brightest Cluster Galaxies (BCGs) have on average shallower density profiles than field galaxies of similar stellar mass \citep[e.g.,][]{Newman13gamma,Newman13b}. The residual scatter left once these scaling relations are taken into account is as small as $6\%$ on the mass density slope \citep{Sonne13}, highlighting once more the high degree of self-similarity across the population of early-type galaxies.

The origin of such a regularity is not fully understood. If massive, central early-type galaxies are predominantly grown by mergers \citep[][]{Naab09,ShankarPhire,ShankarRe,Bai14,Shankar15,BuchanShankar,Lidman16},
stochastic processes may tend to alter pre-existing scaling relations, or at least increase the scatter in the population \citep[e.g.,][]{Nipoti09a,Nipoti09b,Shankar13}, unless additional mechanisms are present
such as gas dissipation \citep[e.g.,][]{Robertson06,Oser10,Remus13,Sonne14merge,Remus16}. More recent numerical and semi-analytic studies suggest that even collisionless mergers could create tight correlations close to those observed \citep[e.g.,][]{Shankar14,Taranu13,Taranu15}.
\citet{Remus13} specifically investigated the evolution of the total mass density slope in a set of galaxies in a cosmological simulation, and were able to naturally produce slopes close to isothermal consistent with observations. Despite recent progress \citep[e.g.,][and references therein]{Dubois13,Remus13}, however, the still unclear sub-grid physics and large computational power required to run high-resolution cosmological simulations do not allow for a proper quantitative comparison between detailed models and observations.

A powerful, complementary approach to cosmological simulations and semi-analytic models is to use semi-empirical models \citep[e.g.,][]{DuttonTreu}.
The latter, by design, are an extremely effective way of making use of the known properties of galaxies together with minimal theoretical inputs to make testable predictions on a set of observables, and set unique, independent constraints on galaxy evolution processes.

In this work we use the observed correlations between photometric properties of massive galaxies (i.e., effective radius, stellar mass), together with basic assumptions on their host dark matter halos, as inputs to create large mock galaxy catalogs.
We then test the validity of the (few) input assumptions and parameters, by comparison with data at different environments
spanning from the field to the cluster halo mass scale. In the following papers of this series we will specifically include velocity dispersion in our mocks as an additional probe for models, and study the evolution of the mass density profile with redshift.

The structure of this work is the following.
In \sect\ref{sec|Method} we describe the method used to construct mock samples of galaxies and relative measurements of the density slope and dark matter fraction.
In \sect\ref{sec|DependsOnGalaxies} we present the predictions for the dependence of the total mass density profile on galactic properties,
and environment in \sect\ref{sec|MassProfileDependenceEnvironment}.
In \sect\ref{sec|Discu} we discuss our results and conclude in \sect\ref{sec|Conclu}.

In the following we will adopt the reference cosmology
with parameters
$\Omega_{\rm m}=0.30$, $\Omega_{\rm b}=0.045$, $h=0.70$,
$\Omega_\Lambda=0.70$, $n=1$, and $\sigma_8=0.8$, to match those usually
adopted in the observational studies on the stellar
mass function and lensing considered in this work. We will also by default adopt
a Salpeter \citep{SalpeterIMF} as our reference Initial Mass Function (IMF),
though we will carefully discuss the implications on our model predictions when switching to a Chabrier \citep{Chabrier03} or even variable \citep{Cappellari15} IMF.

\section{Method}
\label{sec|Method}

\subsection{The Semi-Empirical Model}
\label{subsec|EmpiricalModel}

The aim of this work is to compare semi-empirical models based on $\Lambda$CDM halo relations coupled to observed galaxy scaling relations,
to mass modeling analysis principally from strong lensing.
Our semi-empirical approach makes use of mostly observational quantities with only a few,
basic theoretical inputs. In this respect, it is extremely powerful
as it does not rely on any physical assumption for evolving galaxies, e.g., via mergers and/or in-situ star formation.

Our procedure to build mock galaxies relies on the following steps:
\begin{itemize}
  \item We take large catalogues of central, early-type galaxies with measured stellar masses, projected effective radii, and S\'{e}rsic indices \citep{Sersic63}.
  \item To each galaxy we assign a host dark matter halo via abundance matching relations.
  \item We associate to each central galaxy a 3D \devac\ \citep{deVac} or S\'{e}rsic stellar profile \citep{Sersic63}, according to which data set we compare to.
  Each profile depends on its specific projected effective radius and S\'{e}rsic index (equal to four for a \devac\ profile),
  as detailed in \app\ref{app|Sersic3Dprofile} \citep[e.g.,][]{Lima99}.
  \item A \citet[][NFW hereafter]{NFW} or Einasto \citep{Einasto65} profile characterizes instead each host dark matter halo.
  \item For each galaxy we then build the scale-dependent, total mass density profile.
\end{itemize}
Although we will adopt the above as our reference model, we will also extensively discuss the impact
on our predictions induced by relaxing one or more of the previous assumptions.
We will consider changes in the both the dark matter and stellar profiles, in the IMF, in the stellar mass-halo mass mapping, and also in dark matter concentration.

In detail, for our mocks we extract galaxies from the \citet[][]{Meert15} sample derived from the Sloan Digital Sky Survey (SDSS) DR7 spectroscopic sample \citep{2009ApJS..182..543A}
in the redshift range $0.05<z<0.2$, and with a probability $P(E)>0.85$ of being elliptical galaxies based on the Bayesian automated morphological classifier by \citet{Huertas11}.
The vast majority of the early-type galaxies in this sample are labeled as ``central'' galaxies according to the \citet{Yang07} host halo catalog.
Stellar masses are obtained from integrated light profiles multiplied by the color-dependent mass-to-light ratios of \citet{Bell03SEDs} renormalized to a Chabrier IMF.
In the following, depending on the data set to compare with, we will consider
three different light profiles, a \devac+Exponential (the ``\texttt{cmodel}'', see \citealt[][]{Bernardi10}),
a S\'{e}rsic+Exponential, and pure S\'{e}rsic profiles \citep[see, e.g.,][]{Bernardi13}.
Making use of directly measured physical quantities to build the mock has the advantage that no assumptions have
to be made about, e.g., intrinsic correlations and scatters (e.g., Gaussian) among galaxy sizes, stellar masses and S\'{e}rsic indices.
It is important to stress that SDSS is characterized by an average seeing of $\sim 1.5$ arcsec which would correspond to projected scales of $R\sim 3$ kpc at the average
redshift of $z\sim 0.1$. This observational limit is small enough not to bias the measurement of the stellar profile on scales $R\gtrsim 0.2\, R_{\rm e}$ for galaxies
$\mstare \gtrsim 4\times 10^{11}\, \msune$, though it mostly prevents the possible detection of a stellar ``core'' or flattening below this scale. In the next Sections
we will discuss if extrapolations of a power-law S\'{e}rsic profile to very small scales is supported by strong lensing and kinematic data.

On the dark matter side, our reference will be the NFW profile with mass density
\textbf{\begin{equation}
\rho(r)=\frac{\rho_0}{\left(r/r_s\right)\left(1+r/r_s\right)^2} \, ,
\label{eq|NFW}
\end{equation}}
where $r_s$ is the scale radius, and $\rho_0$ is a reference density. Where relevant,
we will also consider other analytic forms for the dark matter profile, namely a ``cored'' NFW profile \citep{Newman13gamma}
\textbf{\begin{equation}
\rho(r)=\frac{b\rho_0}{\left(1+br/r_s\right)\left(1+r/r_s\right)^2} \, ,
\label{eq|coredNFW}
\end{equation}}
with a reference value of $r_{\rm core}=r_s/b=14$ kpc as the one adopted by \citet{Newman13gamma}, and an Einasto profile, expressed as \citep[see, e.g.,][their Appendix A]{Mamon05}
\textbf{\begin{equation}
\rho(r)=\rho_{-2}\exp{(2\mu)}\exp\left[-2\mu \left(\frac{r}{r_{-2}}\right)^{1/\mu}\right] \, .
\label{eq|Einasto}
\end{equation}}
In \eq\ref{eq|Einasto} $\rho_{-2}$ is the local mass density at $r_{-2}$, the radius at which the logarithmic
slope of the density profile is equal to -2, which is equal to $r_s$ for a NFW profile. We will further discuss below the values of $\mu$ adopted in this work.

Following several previous attempts in the literature \citep[e.g.,][]{DuttonTreu,Oguri14},
we also consider NFW profiles modified by the inclusion of adiabatic contraction and/or expansion.
As in \citet{Dutton07}, assuming spherical collapse without shell crossing, we express
the actual relation between the final and initial radius
of the mass distribution as $r_f=\Gamma^{\nu}r_i$.
The parameter $\Gamma=r_f/r_i$ is the contraction factor \citep{Blumenthal}, which can be numerically calculated
via the conservation equation
\begin{equation}
M_i(r_i)r_i=M_f(r_f)r_f\, ,
\label{eq|ACmain}
\end{equation}
with $M_i$ the initial (baryonic plus dark matter) mass distribution, and
\begin{equation}
M_f(r_f)=\mstare(r_f)+(1-f_{\rm gal})M_i(r_i)
\label{eq|ACMf}
\end{equation}
the final one, with $f_{\rm gal}$ the ratio between the galaxy stellar (possibly plus gas) mass and the host halo mass \citep[see, e.g.,][]{Barausse12}.
The parameter $\nu$ is set to zero for no profile modification, while it assumes positive or negative values for contraction or expansion,
respectively\footnote{\citet[][and references therein]{DuttonTreu} discuss that $\nu$ may also be correlated
with the shift $\Delta \log \mstare$ from a Chabrier IMF. This effect is rather small and we ignore it in the present study,
though we discuss the effects of changing IMF, host halo masses, etc...}
. It has already been shown that extreme contraction ($\nu \gtrsim 1$) and/or expansion $\nu \lesssim -0.5$
tend to be disfavored by present data \citep{DuttonTreu}. We will therefore limit our analysis to non-maximal effects,
such as the one by \citet{Gnedin04} for contraction ($\nu=0.8$), and a slightly milder expansion with $\nu=-0.3$.

Once a total mass density profile for a galaxy has been constructed,
one direct quantity usually adopted in the literature to compare with lensing data has been $\gamma$, i.e., the \emph{local} logarithmic slope of the spherically averaged total density profile $\rho(r)\propto r^{-\gamma(r)}$. In more recent years, the quantity $\gamma'$ has been more often adopted.
The latter is defined as the mass-weighted slope of the total density profile within a radius $r$, and it is computed as \citep{DuttonTreu}
\begin{eqnarray}
\begin{aligned}
\gamma'(r)&\equiv-\frac{1}{M(<r)}\int_{0}^{r}\frac{d\,{\rm log}\, \rho}{d\,{\rm log}\, x}4\pi x^2\rho(x)dx=\\  
&=-\frac{1}{M(<r)}\left(\rho(r)4\pi r^3-3\int_{0}^{r}\rho(x)4\pi x^2dx\right)=\\  
&=3-{\rm \left. \frac{\emph{d}\,{\rm log} \, M}{\emph{d}\, {\rm log}\, \emph{x}} \right|_{\emph{x}=\emph{r}}}\, .
\label{eq|gamma}
\end{aligned} 
\end{eqnarray}
The mass-weighted slope $\gamma'(r)$, specifically computed within the effective radius \re, has been shown to well approximate the
slope measured in joint strong lensing and stellar kinematics studies \citep{Sonne13}.
For power-law density profiles, $\gamma'=\gamma$. In the following, unless otherwise noted, we will always refer to the mass-weighted slope $\gamma'$ computed at \re. \citet[][cfr. their \figu6]{Sonne13} and \citet{DuttonTreu} have shown that the mass-weighted $\gamma'$ computed as in \eq\ref{eq|gamma} is a good proxy of $\gamma'$ derived from strong lensing and dynamical measurements \citep[see][for a full discussion ]{Sonne13}.

\begin{deluxetable}{ccccc}
\tablewidth{7cm}
\tablecaption{Stellar mass-halo mass relations (\eq\ref{eq|MstarMhalo})}
\tablenum{1}
\tablehead{\colhead{IMF} & \colhead{$\mstare^0$} & \colhead{$\mctwoe^0$} & \colhead{$\alpha$} & \colhead{$\beta$}}
\startdata
Chabrier-SerExp & 10.68 & 11.80 & 2.13 & 1.68 \\
Chabrier-Sersic & 10.71 & 11.82 & 2.13 & 1.66 \\
Chabrier-deVac & 10.63 & 11.80 & 2.17 & 1.77 \\
Salpeter-SerExp & 10.98 & 11.84 & 2.14 & 1.72 \\
Salpeter-Sersic & 11.03 & 11.87 & 2.09 & 1.66 \\
Salpeter-deVac & 10.93 & 11.84 & 2.15 & 1.77 \\
varIMF-SerExp & 10.93 & 11.83 & 2.03 & 1.60 \\
varIMF-Sersic & 10.96 & 11.85 & 2.05 & 1.60 \\
varIMF-deVac & 10.83 & 11.81 & 2.04 & 1.64 \\
\enddata
\tablecomments{Parameters of the relations at $z=0.1$ between the central stellar mass and host halo mass (200 times the critical density) from top to bottom,
for a Chabrier, Salpeter, and variable IMF, and for three different light profiles, a S\'{e}rsic-Exponential, a S\'{e}rsic and a \devac\ profile.
A constant intrinsic scatter of 0.15 dex in stellar mass at fixed halo mass is assumed in all models. For each model the most appropriate stellar
mass function is adopted in the abundance matching routine (\eq\ref{eq|Cum}).}
\label{Table|MstarMhaloRelations}
\end{deluxetable}

\subsection{The stellar mass-halo mass and effective radius-stellar mass relations}
\label{subsec|MstarMhaloReMstar}

In this work, we focus on \emph{central} massive galaxies only, and thus accordingly adopt the mean central stellar
mass-halo mass relation to map galaxies into halos. The complexities behind establishing a secure mapping for satellites have been under intense study for several years
\citep[e.g.,][]{Neistein11,Rodriguez12,WatsonConroy}, and are not vital to the present study focused on massive galaxies with $\log \mstare/\msune \gtrsim 11.5$,
for which the fraction of satellites is very low, at the percent level \citep[e.g.,][]{GuoHong14}.

The empirical stellar mass-halo mass correlations adopted as inputs of our empirical models
are derived from abundance matching between the stellar mass and (sub)halo mass functions
\begin{equation}
\Phi(>\mstare,z)=\Phi_c(>\mhaloe,z)+\Phi_s(>\mhaloe,z)
\label{eq|Cum}
\end{equation}
with $\mhaloe=\mctwoe$ the halo masses defined within $r_{200c}$, such that the average density within $r_{200c}$ is 200 times the critical density of the Universe at redshift $z$.
The $\Phi_c(>\mhaloe,z)$ term refers to the host halo mass function of \citet{Tinker08}, with
the subhalo term $\Phi_s(>\mhaloe,z)$ with mass at infall $\mhaloe$ taken from \citet{Behroozi13}.

For the concentration $c_{200c}=r_{200c}/r_s$ we use the median $c_{200c}-\mhaloe$ relation as derived by \citet{Bene14} for our cosmology,
and include a log-normal scatter of $0.16$ dex.
As we discuss in \app\ref{app|Conc}, the \citet{Bene14}
fit is, at least at $z=0$, in very good agreement with the \citet{Bullock01} model, as revised by \citet{Maccio08} and \citet{DuttonMaccio}.
We find that the exact mass-dependent normalization and scatter of the concentration-mass relation \citep[e.g.,][]{DuttonTreu}, have a relatively small impact to most of our conclusions,
except possibly at cluster scales, as discussed below.
Note that we assume the scatter in halo concentration to be uncorrelated with stellar mass or galaxy size \citep[see discussions in, e.g.,][]{Papa15,Katz16}.
We have anyhow checked that at fixed galaxy stellar mass, size, and host halo mass a systematic variation of $0.16$ dex in halo concentration induces a
relatively small variation of $\sim 4\%$ in the implied $\gamma'$ computed at the effective radius.

The stellar mass function on the left side of \eq\ref{eq|Cum} depends on the assumed light profile. We adopt the results by \citet{Bernardi13},
who provide stellar mass functions for all SDSS galaxies characterized by \texttt{cmodel}, S\'{e}rsic+Exponential, and pure S\'{e}rsic magnitudes (all based on a Chabrier IMF).
As recently discussed by a number of groups, adopting S\'{e}rsic profiles naturally results in more integrated light than \devac\ profiles, and thus, at fixed mass-to-light ratio, larger abundances of massive galaxies
\citep[see, e.g.,][]{Bernardi13,Souza15,Bernardi16,Bernardi16ML,Than16,Bernardi17}. This in turn yields steeper stellar-to-halo mappings, i.e., more massive mean stellar masses at fixed halo mass,
in better agreement with direct measurements of massive brightest cluster galaxies \citep[see, e.g.,][]{Kravtsov14,Shankar14b}.
For each light profile we explore the impact of varying the input IMF, from a Chabrier to a Salpeter and
variable IMF. For the former, we simply add 0.25 dex to each stellar mass \citep[][]{Bernardi10}. For the latter,
we take the velocity dispersion-dependent dynamical/strong lensing mass-to-light ratios from \citet[][their \figu19]{Cappellari16}, expressed as a difference with respect to the stellar (Salpeter) mass-to-light ratio
\begin{eqnarray}
\begin{aligned}
\log \left(M/L\right)_{\rm dyn}-\log \left(M/L\right)_{\rm Salp}=-0.0576+\\
+0.364\times \log \left(\frac{\sigma_e}{200\,\, \rm{km\, s^{-1}}}\right)
\label{eq|MLratioCappellari}
\end{aligned}
\end{eqnarray}
with an intrinsic scatter of 0.11 dex.

The resulting mapping between stellar mass and host halo mass\footnote{Note that \eq\ref{eq|MstarMhalo} is the mean relation for the \emph{entire} galaxy population.
\citet{Dutton10a}, however,
showed that the stellar mass-halo mass relation of early-type galaxies is practically indistinguishable from the one characterizing the total population
for $\mstare \gtrsim 2\times 10^{11}\, \msune$.} is well reproduced
by the following two-power law relation
\begin{equation}
\mstare=\mstare^0 \left(\frac{\mctwoe}{\mctwoe^0}\right)^{\alpha} \left[1+\left(\frac{\mctwoe}{\mctwoe^0}\right)^{\beta}\right]^{-1}\, ,
    \label{eq|MstarMhalo}
\end{equation}
with $\mctwoe=\mhaloe$, valid in the approximate range $10^{10}\lesssim \mstare/\msune \lesssim 10^{12}$ (Chabrier IMF).

\begin{figure*}
    \center{\includegraphics[width=15truecm]{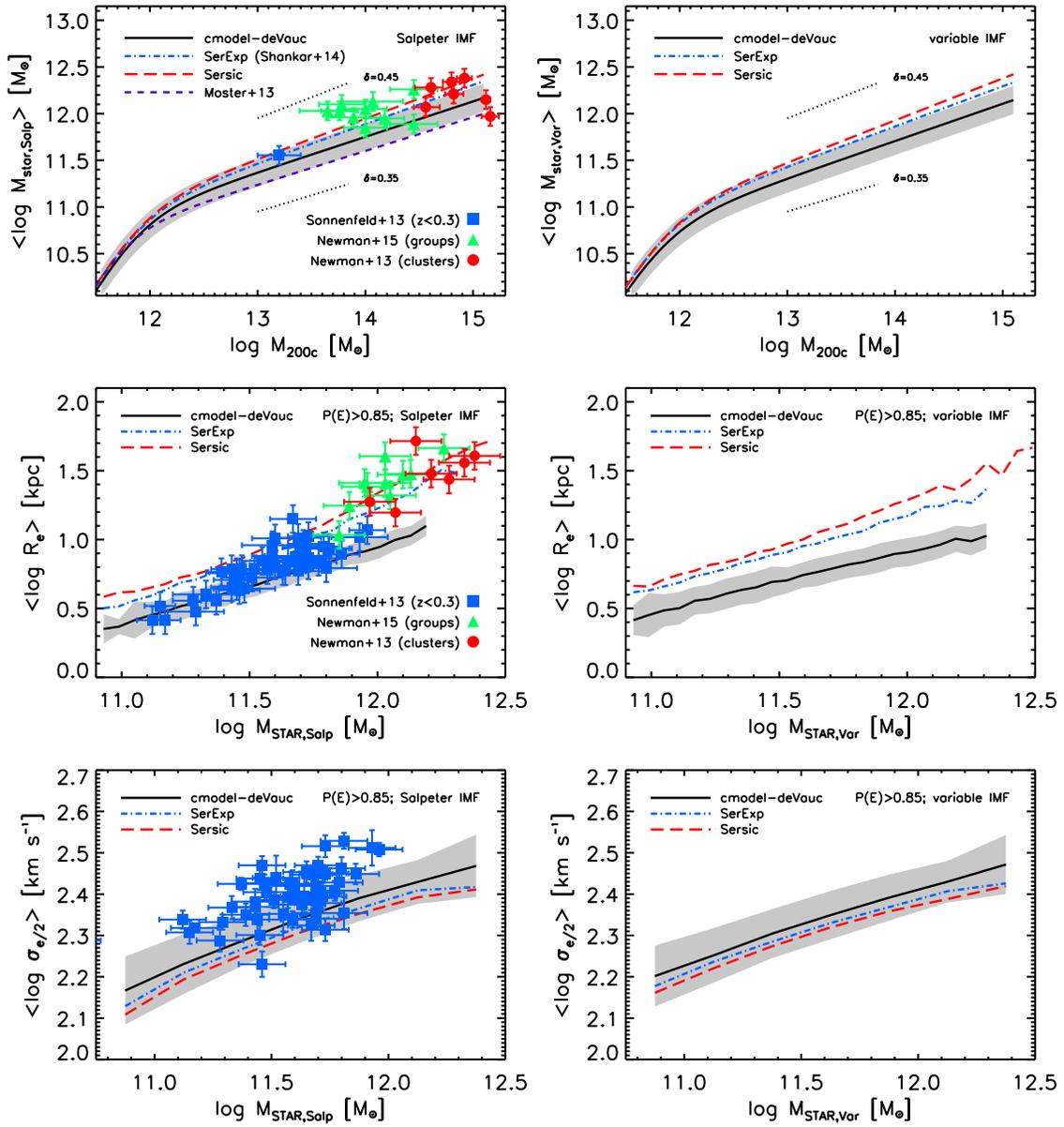}
    \caption{Main galaxy scaling relations adopted in the semi-empirical modeling assuming a Salpeter (\emph{left} panels) or variable (\emph{right} panels) IMF. \emph{Top}: Stellar mass versus halo mass (defined as 200 times the critical density) relation derived from abundance matching between the stellar mass and host halo mass functions assuming an intrinsic scatter of 0.15 dex in $\log \mstare$ at fixed halo mass. The black solid, blue dot-dashed, and red long-dashed lines refer, respectively, to the relations inferred from the \citet{Bernardi13} \texttt{cmodel}, S\'{e}rsic-exponential, and S\'{e}rsic stellar mass functions as described in the text \citep[see also][]{Shankar14b}, while the purple long-dashed line is the relation derived by \citet{Moster13}. \emph{Middle}: Two-dimensional, circularized effective radius  $R_{\rm e}$ versus stellar mass as derived by \citet{Bernardi12} from \texttt{cmodel}, S\'{e}rsic-exponential, and S\'{e}rsic light profiles, as labeled. \emph{Bottom}: Average line-of-sight velocity dispersion within a circular aperture of radius equal to half of the effective radius $\sigma_{\rm e/2}$ as a function of stellar mass for the \texttt{cmodel}, S\'{e}rsic-exponential, and S\'{e}rsic light profiles. In all the left panels we compare, where possible, with the Salpeter-based data by \citet{Sonne13}, \citet{Newman15}, and \citet{Newman13gamma} (blue squares, green triangles, and red circles, respectively). Gray shaded regions in all panels mark the 1$\sigma$ dispersions around the \texttt{cmodel}-\devac\ curves.  \label{fig|MstarMhalo}}}
\end{figure*}

We always assume an intrinsic Gaussian scatter of 0.15 dex in $\log \mstare$ at fixed halo mass, which is a valid approximation especially at high stellar masses and at low redshifts,
in agreement with a number of diverse observational and theoretical studies \citep[e.g.,][and references therein]{GuoHong14,Shankar14b,Gu16,Tinker16}. To include scatter, we follow the methodology outlined in \citet{Shankar14b}. We first fit \eq\ref{eq|MstarMhalo} to the results of abundance matching without scatter, and then vary the parameter $\beta$ until the
implied mock galaxy-halo samples (inclusive of scatter) reproduce well the input stellar mass function within the observational uncertainties.
The final values of the parameters of \eq\ref{eq|MstarMhalo} are given in Table~1 for a Chabrier, a Salpeter,
and a variable IMF, and for each IMF we consider three different light profiles, a \devac, a S\'{e}rsic+Exponential, and S\'{e}rsic profile.
For each combination of light profile and IMF, the implied mean stellar mass-halo mass relation is calculated by inserting in \eq\ref{eq|Cum}
the relevant light profile-dependant stellar mass function \citep{Bernardi13} corrected for an appropriate mass-to-light ratio.

In turn, when assigning a host halo mass to a galaxy in our SDSS sample with measured stellar mass,
we make use of the mean halo mass at fixed stellar mass relation, i.e., the ``inverse'' of \eq\ref{eq|MstarMhalo}.
The latter mapping is computed by first creating a large halo mock catalogue extracted from the halo mass function,
assigning galaxies directly applying \eq\ref{eq|MstarMhalo}, with a scatter of 0.15 dex and parameters given in Table~1, and then self-consistently computing the implied mean and scatter in halo mass as a function of stellar mass. As a consistency check, we make sure that our inverse
relations and related scatters properly reproduce the halo mass function from random galaxy catalogues extracted from the input stellar mass functions.

In \figu\ref{fig|MstarMhalo}, we show the mean stellar mass-halo mass (top), effective radius-stellar mass (middle), and velocity dispersion-stellar mass relation
(bottom) relations assuming a Salpeter (left) or variable IMF (right). As labeled, each panel includes three models corresponding to three different light profiles,
\devac\ (\texttt{cmodel}, black, solid lines with their 1$\sigma$ dispersions marked by gray areas),
S\'{e}rsic-Exponential (blue, dot-dashed lines), and S\'{e}rsic (red, long-dashed lines).

It is clear from the top panels that all our models, and in particular the ones based on S\'{e}rsic profiles, predict mean stellar mass-halo mass relations
with a high-mass end slope of $\delta=\alpha-\beta\sim 0.45$ (see Table~1),
steeper than the \citet{Moster13} (purple, long-dashed line) or \citet{Behroozi13} relations, characterized by slopes $\delta\sim 0.30-0.35$
which would imply up to a factor of $\sim 3$ lower stellar masses at fixed halo mass \citep[see, e.g.,][]{Kravtsov14,Shankar14b,BuchanShankar}.
All our models are in broad agreement with the \citet{Sonne13} data (blue squares), with the S\'{e}rsic-based stellar mass-halo mass relations
particularly well aligned with the \citet{Newman13b} and \citet{Newman15} data (green triangles and red circles, respectively), which are
also characterized by S\'{e}rsic-type light profiles.
All in all, the variable IMF models are characterized by
scaling relations that are very similar to the ones extracted from pure Salpeter models. At best, they might predict stellar masses slightly lower by $\sim 0.1$ dex at low halo masses.
Such a high degree of similarity is expected in the range of high stellar masses considered in this work.

Our models result being significantly steeper than the \citet{Moster13}
relation mainly because ours are based on the \citet{Bernardi13} stellar mass functions, which predict up to two orders of magnitude more massive
galaxies than the \citet{LiWhite} stellar mass function, adopted as the reference one in the \citet{Moster13} analysis. The \citet{LiWhite}
stellar mass function adopted Petrosian magnitudes \citep{Petrosian}, which have long been known
to underestimate the light in the most massive galaxies \citep[e.g.,][]{Bernardi10}. Moreover, the Petrosian magnitudes used by
\citet{LiWhite} were based on the SDSS pipeline, which suffers from bad sky estimates, particularly in crowded fields where
massive galaxies are typically found \citep[see, e.g.,][and references therein]{Bernardi10,Bernardi12}. More recent work
by \citet{Souza15} and \citet{Bernardi16ML} confirms that the \citet{LiWhite} photometry was indeed biased faint.
Finally, \citet{LiWhite} adopted stellar mass-to-light ratios based on templates
from \citet{BlantonRoweis}, which are inappropriate for massive galaxies \citep[e.g.,][]{Bernardi13,Bernardi16,Bernardi16ML,Bernardi17}.
\citet{Bernardi17b} have also carefully demonstrated that the difference with the SDSS pipeline photometry does not depend on whether or not one includes intracluster light, but rather it is truly missing light from the body of the galaxy.

\begin{figure*}
    \center{\includegraphics[width=15truecm]{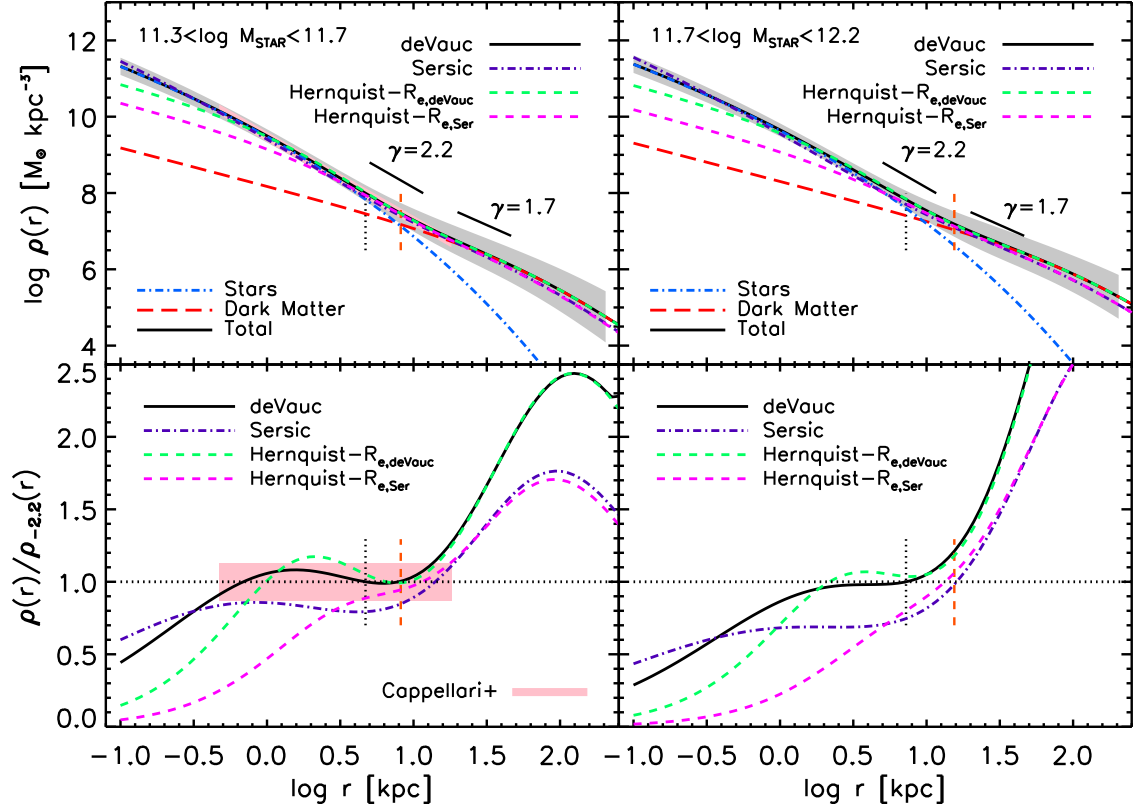}
    \caption{\emph{Top}: Predicted 3D density profiles for the combined
    galaxy-dark matter halo system, for subsamples of galaxies from the Monte Carlo catalogues with stellar mass in the range $11.3<\log \mstare/\msune<11.7$ (left) and $11.7<\log \mstare/\msune<12.2$ (right). The solid black line is the total density profile predicted for a \devac\ stellar density profile, with the blue dot-dashed, and red long-dashed lines marking the relative contributions of the stellar and dark matter components, respectively. The gray bands mark instead the 2$\sigma$ dispersion in mass density at fixed scale. For comparison, also shown are the mean profiles for S\'{e}rsic and Hernquist profiles (purple, dot-dashed and green, dashed lines, respectively). The pink very thin stripe in the left panel marks the best-fit empirical fit by \citet{Cappellari15} $\rho \propto r^{-2.2}$, strictly valid for galaxies with $10.2<\log \mstare/\msune<11.7$. \emph{Bottom}: Ratios of the (total) mass density profiles and the \citet{Cappellari15} $\rho \propto r^{-2.2}$ fits.
    For lower mass galaxies, the total mass profiles are remarkably consistent with a profile $\rho(r)\propto r^{-\gamma(r)}$ with $\gamma(r) \sim 2.2$, especially around the \devac\ and S\'{e}rsic effective radii (vertical, dotted and dashed lines, respectively).
    \label{fig|LocalRhoMr}}}
\end{figure*}

The middle panels of \figu\ref{fig|MstarMhalo} report instead the two-dimensional effective radius \re\ versus stellar mass for the Salpeter (left) and variable IMF (right), and
for our three reference stellar light profiles, as labeled. It is clear that S\'{e}rsic-based profiles provide significantly more extended light profiles at fixed stellar mass, and thus
proportionally larger effective radii up to a factor of $\gtrsim 2-3$ higher than what inferred from pure \devac\ fits. Within uncertainties, our SDSS-\texttt{cmodel} data compare
reasonably well with the effective radii and stellar masses of the galaxies from the \citet{Sonne13}\footnote{The lenses in \citet{Sonne13} come from the SL2S and SLACS surveys \citep{Auger10,Gavazzi14}.}, which are based on \devac\ profiles. Even closer is the match between our S\'{e}rsic-based size-mass relations with the
data by \citet{Newman13b} and \citet{Newman15}, whose half-light radii and total luminosities are extracted from pseudo-isothermal and S\'{e}rsic profiles, respectively. Note that relatively tight correlations such as those between stellar mass and halo mass, and size versus stellar mass, should naturally induce additional dependencies between galaxies and their dark matter halos. We find, for example, that as a consequence of the relation between galaxy size and stellar mass, and the assumption of a monotonic stellar mass-halo mass relation,
our models produce a roughly linear correlation between effective radius and virial radius, in broad agreement with the one independently measured by \citet{Kravtsov13}.

For completeness, the bottom panels of \figu\ref{fig|MstarMhalo} plot the line-of-sight velocity dispersion within half of the effective radius\footnote{Our SDSS velocity dispersions are defined within an aperture of $R_e/8$ and have been corrected to $R_e/2$ following the mean aperture correction given by \citet{Cappellari06}, based on well-resolved two-dimensional spectroscopic data from SAURON.} as a function of stellar mass, in the same format as for the other panels, for Salpeter (left) and variable (right) IMF and for the three reference light profiles. In this case we find the velocity dispersions in the \citet{Sonne13} data to be slightly higher by  $\sim 0.05$ dex, at fixed stellar mass or, alternatively, stellar masses to be lower by $\sim 0.1$ dex at fixed velocity dispersion. Irrespective of this, it is indeed quite remarkable that, despite possible differences and biases in the selections, and in the surface brightness profile fitting and/or background subtraction algorithms, the data by \citet[][]{Sonne13}, \citet{Newman13b} and \citet{Newman15} show with respect to our SDSS catalogs offsets of at most $\lesssim 0.1$ dex in effective radius, stellar mass or velocity dispersion.

It could be argued from the apparent (systematic) discrepancy in the (mean) velocity dispersion-stellar mass relations (lower, bottom panel of \figu\ref{fig|MstarMhalo}), that the reference sample of lensed galaxies is biased high in velocity dispersion at fixed stellar mass with respect to SDSS galaxies. We disfavor a substantial bias in the SLACS galaxies as the galaxies do not appear proportionally more compact at fixed stellar mass (middle, left panel in \figu\ref{fig|MstarMhalo}). Indeed, this (relatively small) discrepancy could be mostly ascribed to residual differences between the \texttt{cmodel} and \devac\ light profiles, and/or between mass-to-light ratios \citep[see, e.g.,][and references therein]{Bernardi13,Bernardi17}. \citet{Auger10} claimed a closer match between SLACS galaxies and SDSS, however their comparison was based on the velocity dispersion-stellar mass relation calibrated by \citep{Hyde09a}, who adopted mass-to-light ratios systematically slightly lower than the ones adopted by \citet{Bernardi13}. \citet{Treu06} also did not find any difference between SLACS and SDSS galaxies at fixed velocity dispersion. More relevantly to what follows, we verified that all our main results are robust against the level of mild systematics seen in the left panels of \figu\ref{fig|MstarMhalo}. For instance, we checked that reducing stellar masses in our SDSS mocks by $\sim 0.1$ dex does not affect any of our conclusions. Even when selecting only the SDSS galaxies above the mean \texttt{cmodel} velocity dispersion-stellar mass relation to better match the SLACS sample, also yields similar results. The latter behavior is not unexpected given that, on average, we verified that massive galaxies in all our models share a very weak dependence of \gamm\ on velocity dispersion \citep[see also][]{Poci17,Xu17}.

In the following, we will adopt the scaling relations discussed in this section as inputs for our mock galaxy-halo catalogs. For self-consistency reasons,
we will specifically adopt the \texttt{cmodel} and single-S\'{e}rsic scaling relations when comparing with the \citet{Sonne13} and \citet{Newman13b,Newman15} data, respectively.
Unless otherwise stated, from now on all our stellar masses \mstart\ will refer to Salpeter stellar masses, to make contact with the observations considered in this work, though the input IMF may not necessarily be a Salpeter IMF. In practice, when computing \gamm\ in some models we will assume a different input IMF, Chabrier or variable IMF, but when comparing to the data we will convert our mock stellar masses into the values an observer would measure assuming a Salpeter IMF. This, we remind the reader, corresponds to a positive offset of 0.25 dex  \citep[e.g.,][]{Bernardi10} in stellar mass in case the input IMF is a Chabrier IMF, or a velocity-dependent offset as given in \eq5 for a variable IMF. All other photometric properties, namely effective radii and S\'{e}rsic indices, are instead clearly independent of the choice of input IMF but mainly rely on the choice of assumed light profile.

\begin{figure*}
    \center{\includegraphics[width=15truecm]{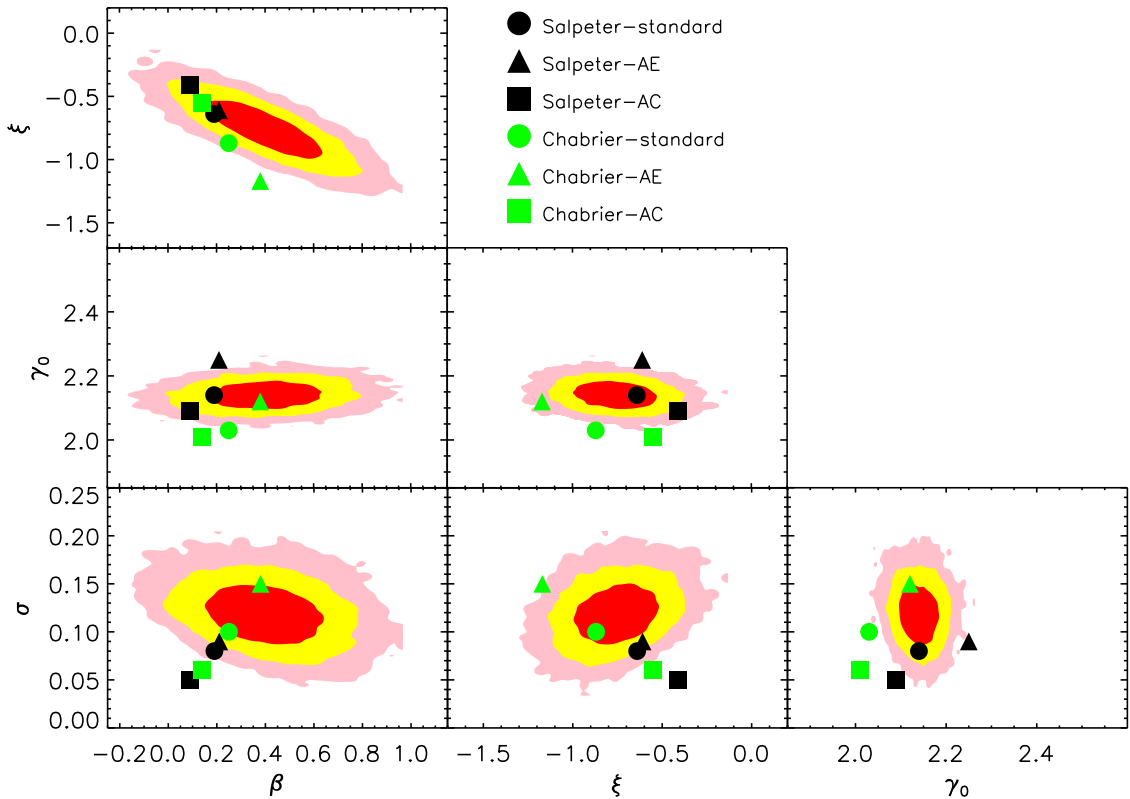}
    \caption{Posterior probability distribution function for the model parameters of \eq\ref{eq|Sonne13gamma}, normalized at $z=0.1$. The red, yellow, and pink regions mark the 1$\sigma$, 2$\sigma$, and 3$\sigma$ contour levels for each pair of the parameter space extracted from the \citet{Sonne13} data. The data are compared with the predictions from the semi-empirical models based with a Salpeter and Chabrier Initial Mass Functions (black and green symbols, respectively). The circles, triangles, and squares refer, respectively, to models with standard profiles, with adiabatic expansion, and with adiabatic contraction, as labeled. \label{fig|GammaPosteriorProbIMF}}}
\end{figure*}

\section{Dependence of \gamm\ on galactic properties}
\label{sec|DependsOnGalaxies}

In this section, we compare the predictions of our models to observational data at $z<0.3$. For simplicity we start by analyzing the average predictions of the semi-empirical models in \sect\ref{subsec|AveragTrend}, and in \sect\ref{subsec|MassProfileDependenceSizeMass} we proceed by analyzing further dependencies on galaxy properties. In this work we will not discuss the dependence of \gamm\ on velocity dispersion, which shares with \gamm\ a strong correlation in the observational errors \citep[e.g.,][]{Sonne13}. We will instead mainly focus on dependencies of \gamm\ on independently-measured quantities such as effective radius and stellar mass. We will get back to velocity dispersions when modeling the velocity dispersion-stellar mass relation in separate work. In \sect\ref{sec|MassProfileDependenceEnvironment} we will discuss environmental trends as quantified by the host halo mass. We expect the latter to be an important variable as massive galaxies at a given stellar mass could live in quite diverse halos.

\subsection{Average trend}
\label{subsec|AveragTrend}

The top panels of \figu\ref{fig|LocalRhoMr} show the predicted 3D mass density profile for galaxies with stellar mass (Salpeter) in the range $11.3<\log \mstare/\msune<11.7$ (left panels) and $11.7<\log \mstare/\msune<12.2$ (right panels). 
The bottom panels of \figu\ref{fig|LocalRhoMr} show the ratio of the total mass density profiles with respect to a pure power-law, a profile of the type $\rho(r)\propto r^{-2.2}$. The latter choice is motivated by the findings of \citet[][see also \citealt{Serra16}]{Cappellari15} who, confirming and extending previous results, infer from two-dimensional stellar kinematics a total mass density profile of $\rho(r)\propto r^{-\gamma}$ with $\langle\gamma\rangle=2.19\pm0.03$, valid in the range $0.1R_{\rm e}$ to $4R_{\rm e}$ and
$10.2\lesssim \log \mstare/\msune\lesssim 11.7$ (their results are reported with pink stripes in the left panels of \figu\ref{fig|LocalRhoMr}).

It is first of all evident from \figu\ref{fig|LocalRhoMr} that the inner regions of massive galaxies are dominated by the stellar component (blue dot-dashed lines), with the dark matter (red long-dashed lines) gradually taking over at larger scales $r \gtrsim \ree$, as already pointed out over a decade ago \citep[e.g.,][]{Borriello03,Mamon05,Mamon05b}.
Nevertheless, the total mean mass density profile (black solid lines) is remarkably well approximated by $\gamma \approx 2.2$ \citep[e.g.,][]{Koop09,Cappellari15}, especially in the region around the effective radius, with a slight tendency towards a somewhat warped profile \citep{Mamon05b,Chae14}. This prediction of the semi-empirical model is in very good agreement with direct strong lensing and dynamical measurements \citep[e.g.,][]{Cappellari15}. It is important to note that the S\'{e}rsic effective radii in massive galaxies tends to become progressively larger than the one based on \devac\ profiles, thus naturally probing regions with flatter density profiles (right panels).

\figu\ref{fig|LocalRhoMr} also includes two Hernquist profiles, with their core radii linearly correlated to the 3D half-mass radii \citep{Hernquist90}. The latter are in turn derived from the 2D circularized effective radii (vertical black dotted and orange dashed lines) adopting \eq\ref{app|R2DR3d}. The green dashed and purple dashed lines show, respectively, the resulting Hernquist profiles for the same stellar masses but different core radii
derived from a \devac\ and S\'{e}rsic 2D effective radius and index (equal to four for the \devac\ profile). It is clear that, irrespective of the input effective radius, the Hernquist profiles are disfavored with respect to a \devac\ or, even better, a pure S\'{e}rsic profile (purple dot-dashed lines), tending to drop too quickly in the inner few kiloparsecs with respect to the data by \citet[][left bottom panel]{Cappellari15}.

\subsection{Dependence on stellar mass and effective radius}
\label{subsec|MassProfileDependenceSizeMass}

While probing the total mass density profile of massive galaxies has been a subject of intense study by a number of teams \citep[e.g.,][]{Koop09,Barnabe11,Ruff11,Bolton12}, only recently the galaxy lensing samples have become large enough to enable the study of its dependence on galactic properties \citep[e.g.,][]{Auger10,Cappellari13b,Chae14,Shu15,Serra16}.
Using strong lensing, \citet{Sonne13} have empirically parameterized the dependence of the mass-weighted \gamm\ on galactic properties as (their Table 4, second column)
\begin{equation}
\langle \gamma' \rangle=\gamma_0+\alpha(z-0.3)+\beta(\log \mstare-11.5)+\xi \log (R_{\rm e}/5) \, ,
\label{eq|Sonne13gamma}
\end{equation}
with $\gamma_0=2.08^{+0.02}_{-0.02}$, $\alpha=-0.31^{+0.09}_{-0.10}$, $\beta=0.40^{+0.16}_{-0.15}$, and $\xi=-0.76^{+0.15}_{-0.15}$, with an average dispersion around the median of
$\sigma_{\gamma'}=0.12^{+0.02}_{-0.02}$.

The data tend to favor a positive increase of \gamm\ with increasing stellar mass, at fixed effective radius. This can be broadly understood in terms of the progressively less significant impact of the (steeper) stellar component when moving to lower mass galaxies, inducing a parallel flattening of the overall profile and thus of \gamm. When fixing the stellar mass and varying the effective radius, \gamm\ is instead expected to steepen with decreasing size, as the profile gets progressively dominated by the steeper stellar component. Similarly, \gamm\ is expected to steepen even faster with mean (stellar) surface density at fixed stellar mass, since $\Sigma = \mstare/(2 \pi R_{\rm e}^{2})$.

In essence, the dependence of \gamm\ on stellar mass and/or effective radius
is connected to probing different portions of the total mass density profile,
from the inner steeper profile of the stellar component to the outer flatter profile of the dark matter component, as well
as varying the relative contributions of the stellar and dark matter components (cfr. \figu\ref{fig|LocalRhoMr}).
Although the trends of \gamm\ summarized in \eq\ref{eq|Sonne13gamma} are broadly expected, their precise measurements
provide invaluable constraints to galaxy evolutionary processes. To this purpose, in this section we will compare
detailed predictions of our semi-empirical models to the \citet{Sonne13} data and specifically to \eq\ref{eq|Sonne13gamma}.

\figu\ref{fig|GammaPosteriorProbIMF} reports the probability distribution function (PDF) measured by \citet{Sonne13} evaluated at the average redshift of our mock SDSS sample, $z=0.1$. The red, yellow, and pink regions bracket, respectively, the 1$\sigma$, 2$\sigma$, and 3$\sigma$ contour levels for each pair of the four parameters $\xi$, $\beta$, $\gamma_0$, and $\sigma_{\gamma'}$. In order to make the closest comparison to the data we treat the outputs of each rendition of our semi-empirical model as in the observations. We apply the same fitting procedure used in \citet{Sonne13} to infer the distribution of \gamm\ across the population of galaxies, which we model as a Gaussian distribution with a mean given by \eq\ref{eq|Sonne13gamma} and with dispersion $\sigma_{\gamma'}$. The fit produces a posterior PDF for the parameters $\gamma_0$, $\beta$, $\xi$ and $\sigma_{\gamma'}$. For each model realization we plot the peak of the posterior PDF in \figu\ref{fig|GammaPosteriorProbIMF}. Due to the large sample size of our mock datasets, statistical uncertainties on these parameters are very small compared to the uncertainty on the same parameters in the \citet{Sonne13} observations, and are therefore omitted in the figure. The circles, triangles, and squares in  \figu\ref{fig|GammaPosteriorProbIMF} refer, respectively, to models with standard profiles, with adiabatic expansion, and with adiabatic contraction, as labeled, computed as discussed in \sect\ref{subsec|EmpiricalModel}.

\begin{figure*}
    \center{\includegraphics[width=15truecm]{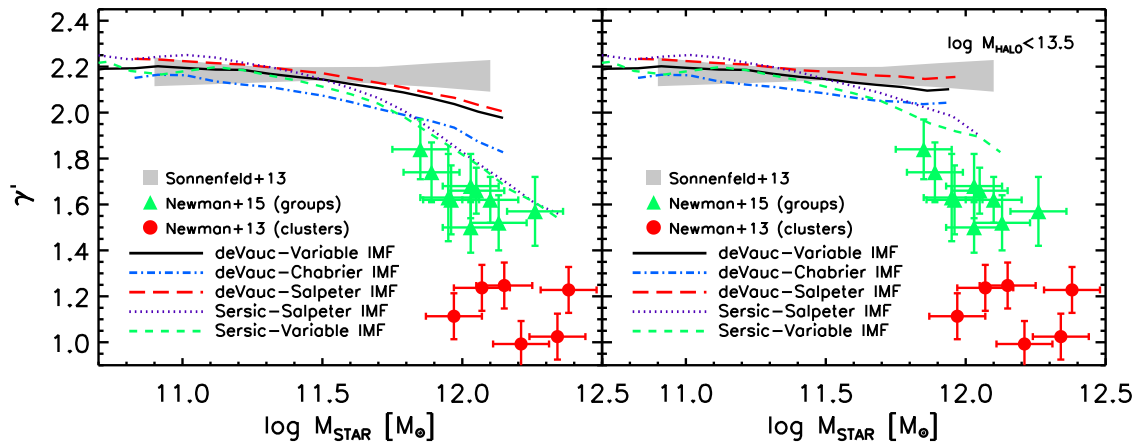}
    \caption{Predicted \gamm\ dependence on (Salpeter) stellar mass for several different models, as labeled, a standard \devac\ profile for a Chabrier, Salpeter and variable IMF (blue dot-dashed, red long-dashed, and black solid lines, respectively), and a S\'{e}rsic profile with a Salpeter and variable IMF (purple dotted, and green dashed lines, respectively). All models assume a NFW profile for the dark matter component. The left and right panels show, respectively, predictions for the same models including all host dark matter halos and restricted to $\log \mhaloe/\msune < 13.5$.
    The data, the same in both panels, are extracted from the \citet{Sonne13} sample (gray band), from the
    ``group'' galaxy sample by \citet[][green triangles]{Newman15}, and from the cluster galaxy sample by \citet[][red circles]{Newman13b}.
    Galaxies of similar stellar mass can show quite different total mass density profiles most probably induced by
    differences in structural properties and in the environments they live in.  \label{fig|GammaMstar}}}
\end{figure*}

The first apparent feature characterizing the Salpeter IMF model is that a standard, uncontracted NFW dark matter profile (black circles) is consistent with the data at the $1-1.5\sigma$ level in all the parameters' spaces identified by \citet{Sonne13}. Adiabatic contraction will instead tend to decrease $\beta$ and increase $\xi$, significantly worsening the match to the data (black squares). The endpoint of adiabatic contraction is in fact to attract dark matter towards more central regions (cfr. \figu\ref{fig|fDM}). By increasing the relative contribution of dark matter at all scales weakens the dependence of \gamm\ on stellar mass, thus effectively decreasing the value of $\beta$ in \eq\ref{eq|Sonne13gamma}. At the same time, adiabatic contraction will also weaken the dependence of \gamm\ on effective radius, rendering $\xi$ less negative, as the profile tends to steepen at larger scales increasing \gamm. Conversely, adiabatic expansion (black triangles) tends to progressively lower the dark matter contribution in the inner regions allowing for a more dominant role of the steeper stellar component, thus increasing the zero point $\gamma_0$, increasing $\beta$, and decreasing $\xi$. \figu\ref{fig|GammaPosteriorProbIMF} shows that even milder adiabatic contractions or expansions, as the ones adopted here, tend to be ruled out at the $\gtrsim 3\sigma$ level in at least two or more pairs of parameters. Adiabatic expansion, in particular, appears to be excluded at high significance (black triangles).

The second important point to mention is that a Chabrier IMF (green symbols in \figu\ref{fig|GammaPosteriorProbIMF}) is also disfavored by the data, irrespective of any adiabatic contraction/expansion included in the models. The mean $\gamma_0$, in particular, appears to be $>3\sigma$ away the values constrained by the data, irrespective of the exact dark matter profile assumed (green symbols in the middle panels). The model with a Chabrier IMF predicts in fact lower stellar masses at fixed halo mass, which in turn reduces the contribution of the steeper stellar component and inducing a flatter total mass density profile (see \figu\ref{fig|LocalRhoMr}), especially around the effective radius, in tension with the strong lensing data by \citet[][middle panels in \figu\ref{fig|GammaPosteriorProbIMF}]{Sonne13}.

In Appendix~\ref{app|OtherModels} we also show a series of other systematic variations in the inputs of our reference semi-empirical model (\sect\ref{subsec|EmpiricalModel}). First off, the PDFs for a Salpeter and variable IMF are quite similar and thus they are both consistent with the \citet{Sonne13} strong lensing data (\figu\ref{fig|App:GammaPosteriorProbIMFvar}). This is partly expected given that the input scaling relations appear to be quite similar for both choices of IMF (\figu\ref{fig|MstarMhalo}). In future work we will explore additional observational proxies such as the velocity dispersion-stellar mass relation to hopefully break these type of degeneracies. Assuming a Hernquist stellar profile (\figu\ref{fig|App:GammaPosteriorProbHern}) also has a relatively mild effect on the model predictions, as expected given that it mainly flattens out the mass density profile within the regions inside the effective radius (\figu\ref{fig|App:SersicProfileExample}), leaving the total mass density profile almost unaltered around $R_{\rm e}$. Nevertheless, the model characterized by a Hernquist profile in the stellar component tends to be only marginally consistent with the data at the $\sim 2-3\sigma$ level. Moreover, a shallower stellar profile in the inner regions of intermediate-mass early-type galaxies tends to be in tension with strong lensing and 2D dynamical measurements which point to steep total mass density profiles \citep[e.g.,][]{Auger10,Cappellari15}, as we showed in \figu\ref{fig|LocalRhoMr}. Introducing a core in the dark matter (\figu\ref{fig|App:GammaPosteriorProbCore}) severely worsens the match with the data in all parameters, mainly inducing steeper mass densities profiles thus larger \gamm\ values (middle panels). In this case adiabatic contraction may help bringing model predictions closer to the data, though discrepancies remain at the $\sim 3\sigma$ level, especially in terms of the very weak dependence on stellar mass ($\beta \sim 0$) and also significantly weaker in size (green squares). As expected, also models characterized by combinations of Hernquist plus cored dark matter profiles tend to be significantly less aligned with the data, being excluded at the $\sim 3\sigma$ level, irrespective of any adiabatic contraction or expansion (\figu\ref{fig|App:GammaPosteriorProbCoreHern}). Switching to the \citet{Moster13} stellar mass-halo mass relation (\figu\ref{fig|App:GammaPosteriorProbMoster}) has a relatively small impact to model predictions, as the \citet{Sonne13} data are dominated by intermediate-mass galaxies, and not brightest cluster galaxies, where the difference becomes most prominent (\figu\ref{fig|MstarMhalo}). Nevertheless the adiabatic contraction or expansion tend to still be excluded at $\sim 3\sigma$ level.

\begin{figure*}
    \center{\includegraphics[width=18truecm]{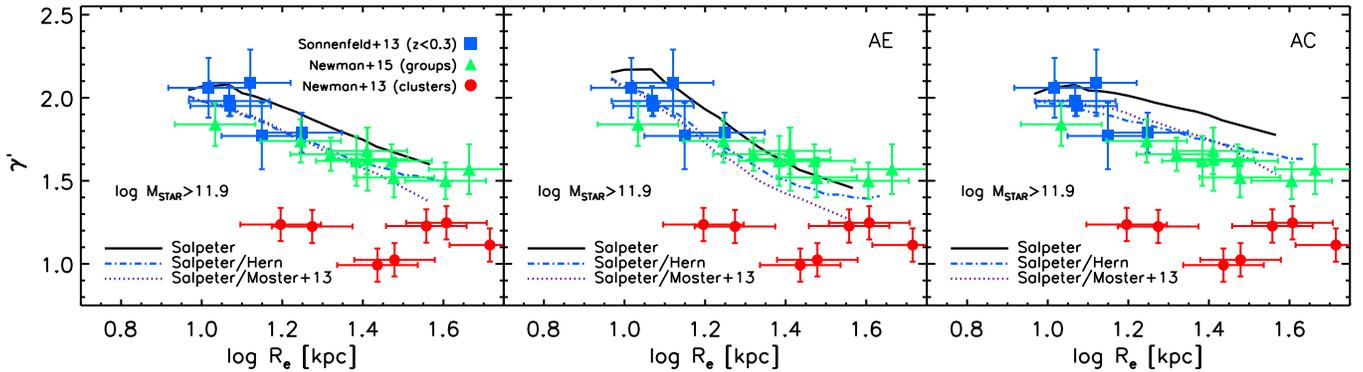}
    \caption{Predicted \gamm\ dependence on effective radius for galaxies with $\log \mstare/\msune>11.9$ for three different models, as labeled: a standard S\'{e}rsic+NFW (black, solid lines), a Hernquist+NFW (blue, dot-dashed lines), and a S\'{e}rsic+NFW adopting the stellar mass-halo mass relation by
    \citet[][purple, dotted lines]{Moster13}. All models assume a Salpeter IMF. The left, middle, and right panels show predictions for the same models assuming dark matter halo profiles uncontracted, with Adiabatic Expansion, and Adiabatic Contraction, respectively.
    Filled, blue squares, green triangles, and red circles are data from the $z<0.3$ \citet{Sonne13} sample, from the ``group'' galaxy sample by \citet{Newman15}, and from cluster galaxy sample by \citet{Newman13b}, all with $\log \mstare/\msune>11.9$ and Salpeter IMF.
    Galaxies of similar stellar mass and effective radius can still show quite different total mass density profiles most probably induced by the different environments they live in. \label{fig|GammaReCluster}}}
\end{figure*}

\section{Dependence of \gamm\ on Environment}
\label{sec|MassProfileDependenceEnvironment}

So far, we have been discussing trends of \gamm\ with galaxy properties, ignoring the possible effects of the larger-scale environment, as labeled by the host dark matter halo mass. In order to add this important dimension to our analysis we need to extend our data sets adopted as comparison. As discussed by \citet[][]{Treu09}, the SLACS survey, which is the foundation of the \citet{Sonne13} sample, is dominated by massive galaxies not in cluster-scale halos\footnote{We point out that when comparing with \citet{Sonne13} in \figu\ref{fig|GammaPosteriorProbIMF} or in the Appendices, we do not apply any halo mass cut in our mock galaxies, though we verified that retaining only galaxies with, say, $\log \mhaloe/\msune<13.5$, does not alter our main results.} (only $\sim 17\%$ of the original sample resides at the center of massive groups and clusters). A stacked weak lensing analysis of a subset of those galaxies also indicated a mean
host halo mass of $\mhaloe \sim 1.6\times 10^{13}\, \msune$ \citep{Gavazzi07}, with an intrinsic scatter of a factor of $\sim 2$ \citep{Newman15}.
The \citet{Sonne13} sample also tends to run out of galaxies above $\mstare \sim 9\times 10^{11}\, \msune$. We thus complemented our reference observational sample with 10 galaxies from \citet{Newman15}, specifically selected to be very massive galaxies with mean $\mstare \sim 10^{12}\, \msune$ (Salpeter IMF) at the center of massive groups with an average mass $\mhaloe \sim 10^{14}\, \msune$, and seven very massive galaxies by \citet{Newman13b}, all at the center of clusters and with stellar mass $\mstare \gtrsim 10^{12}\, \msune$ within clusters of mass $4\times 10^{14}\lesssim \mhaloe/\msune \lesssim 2\times 10^{15}$, with total mass density profiles obtained via a combination of weak and strong lensing, resolved stellar kinematics within the BCG and X-ray kinematics (Table~9 in \citealt{Newman15}). All halo masses are always self-consistently defined as $\mctwoe$ (see \sect\ref{sec|Method}).

\subsection{Dependence on host halo mass}
\label{subsec|MassProfileDependenceEnvironment}

To start off with, we compare in \figu\ref{fig|GammaMstar} the dependence of \gamm\ versus (Salpeter) stellar mass for the whole \citet{Sonne13} lensing sample (gray band), with the ``group'' galaxy sample by \citet[][green triangles]{Newman15}, and with the cluster galaxy sample by \citet[][red circles]{Newman13b}. The \citet{Sonne13} data indicate a very weak dependence of \gamm\ on stellar mass. In particular, at masses around $\mstare \sim 10^{12}\, \msune$, the three samples predict very different total mass density slopes, ranging from values of \gamm$\sim 2.2$ to values as low as \gamm$\sim 1.1-1.2$. To better understand the origin of such huge discrepancies in these different data sets, we compare with the predicted mean \gamm\ for several different models, a standard \devac\ profile for a Chabrier, Salpeter and variable IMF (blue dot-dashed, red long-dashed, and black solid lines, respectively), and a S\'{e}rsic profile with a Salpeter and variable IMF (purple dotted, and green dashed lines, respectively), as labeled. The first point to note is that all models predict a more or less rapid drop of \gamm\ with increasing stellar mass, irrespective of the input stellar profile or IMF (left panel). However, if in our mocks we select only ``field'' galaxies (right panel), i.e. we disregard galaxies at the center of massive groups and clusters above, say, $\log \mhaloe=13.5$ (which is around the upper limit probed by \citet{Sonne13}), we tend to find steeper profiles in high stellar mass galaxies induced by the lower contribution of the host dark matter halo characterized by flatter profiles. This simple cut in host halo mass is sufficient to reconcile \devac\ model predictions with the \citet{Sonne13} data (gray band). It can be seen that both the \devac\ models with Salpeter and variable IMF match observations, with the Chabrier IMF model predicting generally lower \gamm, in line with what discussed in reference to \figus\ref{fig|GammaPosteriorProbIMF} and \ref{fig|App:GammaPosteriorProbIMFvar}.

\begin{figure*}
    \center{\includegraphics[width=18truecm]{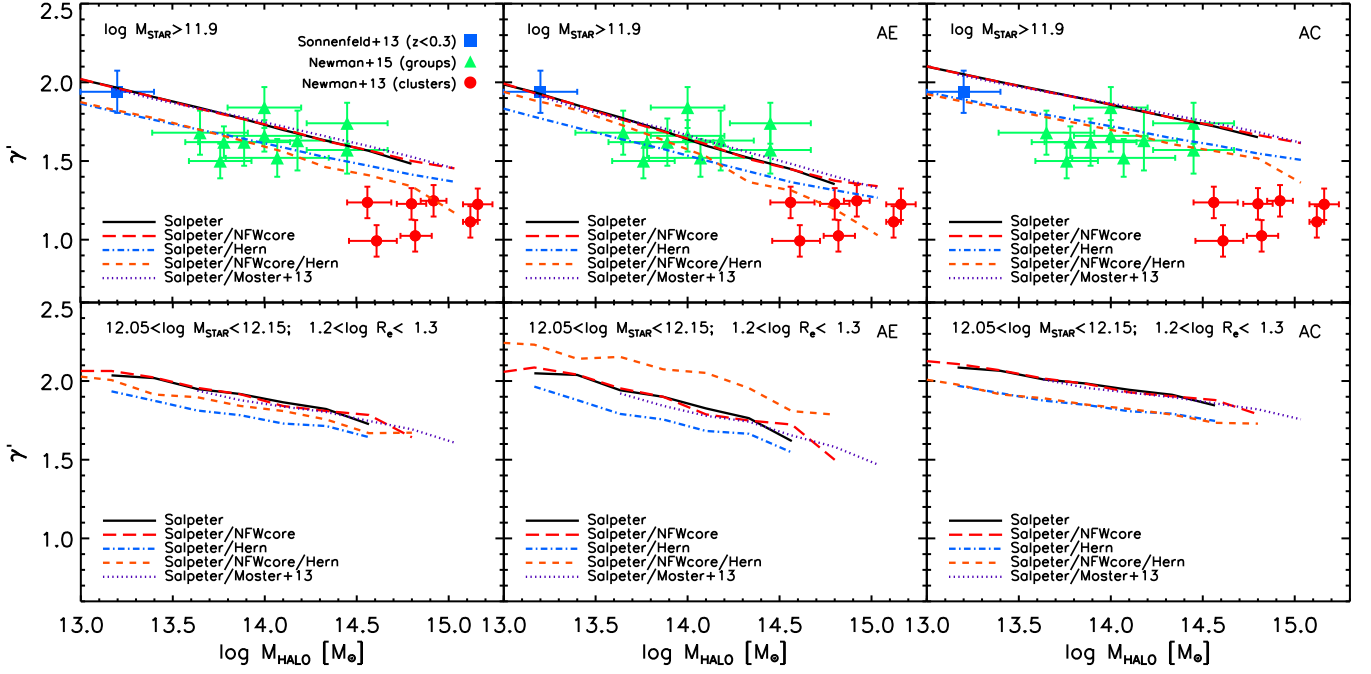}
    \caption{Predicted dependence of \gamm\ on host halo mass for five different models,
    a standard S\'{e}rsic+NFW (black, solid lines), a S\'{e}rsic+cored NFW (red, long-dashed lines), Hernquist+NFW (blue, dot-dashed lines), a Hernquist+cored NFW (orange, dashed lines), and a S\'{e}rsic+NFW adopting the stellar mass-halo mass relation by
    \citet[][purple, dotted lines]{Moster13}. All models assume a Salpeter IMF. The left, middle, and right panels show predictions for the same models assuming dark matter halo profiles uncontracted, with Adiabatic Expansion, and Adiabatic Contraction, respectively.
    As in \figu\ref{fig|GammaReCluster}, filled, blue squares, green triangles, and red circles in the top panels are data from \citet{Sonne13}, \citet{Newman15}, and \citet{Newman13b}. The bottom panels show the predicted trends for a narrow bin in stellar mass and effective radius, as labeled.
    The predicted total mass density profile has a relatively strong dependence on halo mass with \gamm$\propto \mhaloe^{0.2-0.3}$, mostly because a higher host halo mass induces an overall flatter mass density profile and thus lower \gamm. \label{fig|GammaDependenceHalo}}}
\end{figure*}

However, even after excluding massive host halos, the S\'{e}rsic-based models tend to fall to lower \gamm\ values at large stellar masses.
This key difference with respect to the predictions of \devac-based models could be mostly ascribed to the progressively larger effective radii in more massive galaxies with S\'{e}rsic profiles (middle panels of \figu\ref{fig|MstarMhalo}). Having larger radii, the S\'{e}rsic-based models probe larger and flatter portions of the mass density profile, in which the dark matter starts dominating, as emphasized in the right panel of \figu\ref{fig|LocalRhoMr}. We conclude that the Newman et al. galaxies show significantly flatter total mass density profiles with respect to Sonnenfeld et al. mainly because they are central galaxies in groups and clusters, thus they are \emph{physically} more extended and better described by S\'{e}rsic profiles \citep[][]{Bernardi11a,Bernardi11b,Bernardi12}, and also characterized by significantly more massive host dark matter halos.

To further probe the dependence on structural properties, in \figu\ref{fig|GammaReCluster} we plot \gamm\ against effective radius for the most massive galaxies in the \citet[][]{Sonne13} catalog, and the group and cluster galaxy samples of \citet{Newman15} and \citet{Newman13b}, shown by blue squares, green triangles, and red circles, respectively. We include three semi-empirical models in this figure, characterized by a standard stellar S\'{e}rsic and dark matter NFW profiles (black, solid lines), a stellar Hernquist and dark matter NFW profiles (blue, dot-dashed lines), and a stellar S\'{e}rsic and dark matter NFW based on the stellar mass-halo mass relation by
\citet[][purple, dotted lines]{Moster13}. For consistency with the data, in \figu\ref{fig|GammaReCluster} (as well as in \figu\ref{fig|GammaDependenceHalo}) we show results for the subsample of mock galaxies with $\log \mstare >11.9$, which is comparable to the lower limit in stellar mass probed in the Newman et al. samples, and still low enough to retain a statistically relevant number of galaxies in our mocks. For the same reason, we also restrict the \citet{Sonne13} data to only galaxies with $\log \mstare >11.9$. Despite the variety of mass profiles and stellar-dark matter mappings the predicted \gamm\ distributions show quite similar behaviors, differing only in normalization by just a modest amount of $\lesssim 10\%$. More interestingly, the predicted mean \gamm\ substantially decreases with increasing size, as the (flatter) dark matter component becomes more dominant at larger scales (see discussion in \sect\ref{subsec|MassProfileDependenceSizeMass}). Our model predictions, at least the ones without adiabatic contraction, are in fact in broad agreement with the data, and the predicted trend is by itself sufficient to explain a large portion of the difference between the \gamm\ values in the \citet{Sonne13} sample and in the group galaxies of \citet{Newman15}.

However, the models appear to still be highly inconsistent with the \gamm\ measurements by \citet{Newman13b} for BCGs (red circles). Their median \gamm\ lays significantly below our estimates, and even below the data by \cite{Newman15} for galaxies having similar stellar masses and effective radii but residing in group-scale halos. The mean offset in stellar mass between the
\citet{Newman15} group sample and the \citet{Newman13b} cluster sample is in fact only $\sim 0.17$ dex, which is too modest to explain the strong observed systematic difference in \gamm\ purely in terms of stellar mass. This finding is consistent with the hypothesis that halo mass may play a non negligible role in shaping the overall mass density profiles of very massive galaxies at the center of clusters \citep[see also][]{Koch01}.

To explore this issue further, we present in \figu\ref{fig|GammaDependenceHalo} the predicted dependence of \gamm\ on host halo mass from our reference models, compared with data from \citet{Sonne13} for field/group galaxies (blue squares), \citet{Newman15} for group central galaxies (green triangles), and \citet{Newman13b} for BCGs (red circles). It is first of all apparent, as also pointed out by \citet{Newman15}, that \gamm\ depends on environment/halo mass. While some of the data appear more mixed in the \gamm-$R_{\rm e}$ plane (\figu\ref{fig|GammaReCluster}), they tend to separate well when \gamm\ is expressed as a function of host halo mass. Predictions for a series of models (with Salpeter IMF) are included in \figu\ref{fig|GammaDependenceHalo}: a standard stellar S\'{e}rsic and dark matter NFW profiles (black, solid lines), a stellar S\'{e}rsic and dark matter cored NFW profiles (red, long-dashed lines), a stellar Hernquist and dark matter NFW profiles (blue, dot-dashed lines), a stellar Hernquist and cored NFW profiles (orange, dashed lines), and a stellar S\'{e}rsic and dark matter NFW profiles adopting the stellar mass-halo mass relation by \citet[][purple, dotted lines]{Moster13}.

\begin{figure*}
    \center{\includegraphics[width=17truecm]{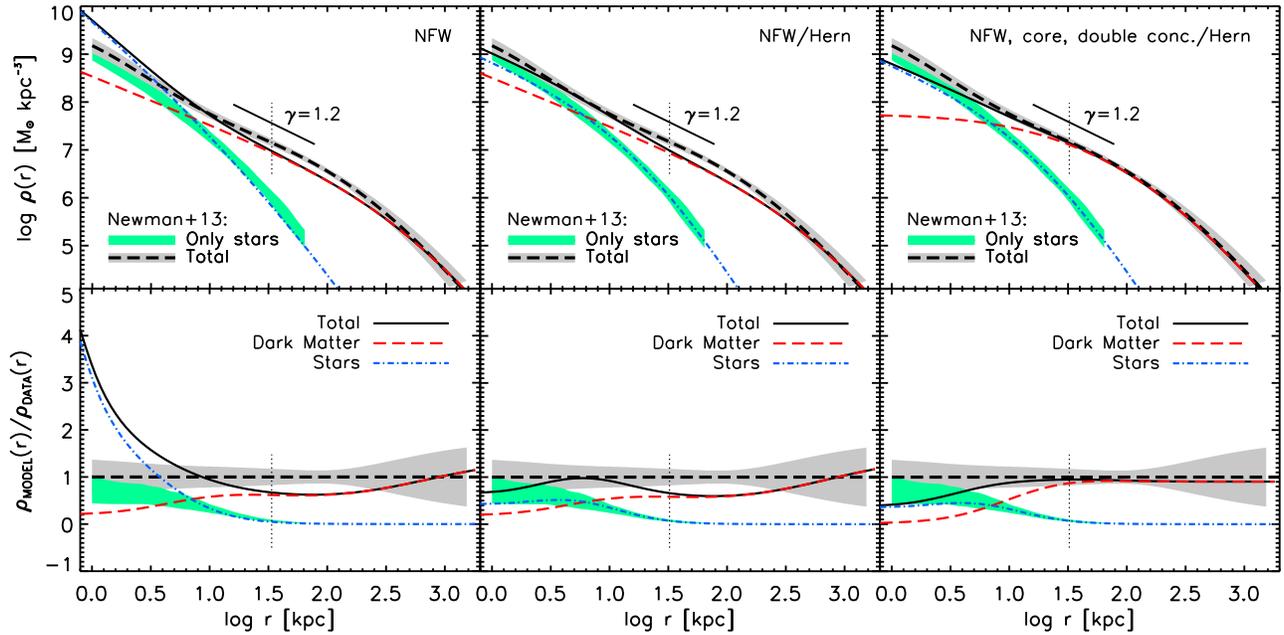}
    \caption{Total mass density profile predicted from three models, the reference S\'{e}rsic+NFW (left), Hernquist+NFW (middle), and a Hernquist+cored NFW  with $r_{\rm core}=14$ kpc and a mean concentration higher by factor of two (right). The top panels compare the model outputs with the data by \citet{Newman13b}. In each panel the black long-dashed line with gray area mark, respectively, the mean profile and its 1$\sigma$ dispersion, while the light green area roughly brackets the 1$\sigma$ dispersion in the average stellar profiles of the \citet{Newman13b} central galaxies. The bottom panels show the ratios between model predictions and data. The stellar and dark matter contributions to the total mass density profiles are reported by blue triple dot-dashed, and red long-dashed lines, respectively. An inner flatter stellar component and higher dark matter concentrations tend to yield better fits to the data. The vertical dotted lines in all panels mark the position of the mean effective radius in the mock galaxies. \label{fig|GammaInClusters}}}
\end{figure*}

All models, especially the ones with no adiabatic contraction, predict a drop of \gamm\ with increasing halo mass in line with the data.
Our reference model with uncontracted NFW dark matter profiles, in particular, predicts (left panel) a dependence on halo mass of the type
\begin{equation}
\gamme\ \propto {\mhaloe}^{p}
\label{eq|gammaMhalo}
\end{equation}
with $p=-0.30\pm0.01$, in good agreement with \citet{Newman15} who find $p=-0.33\pm0.07$ when combining different samples of lensed galaxies in the field, in groups, and in clusters (each single data set, by itself, does not show any dependence in halo mass). The value of $p$ we find is strictly valid for $\log \mhaloe/\msune \gtrsim 13$, with a possible gradual flattening below this mass scale. To make a fairer comparison with the data, we again consider in \figu\ref{fig|GammaDependenceHalo} only galaxies from our semi-empirical models with $\log \mstare/\msune>11.9$. However, we have verified that the dependence of \gamm\ on halo mass is quite independent of the exact cuts in stellar mass or effective radius. As such it could be considered as an addition to \eq6. We note, however, that observational calibrations of the slope $p$ require uniform and precise measurements of strong and weak lensing galaxies over a large range of host halo masses. Non-negligible errors in halo mass could in fact wash out some of the trends with halo mass \citep[e.g.,][]{Shankar14}. The dependence of \gamm\ on halo mass is a genuine effect and only partly induced by the increase of effective radius with stellar mass. This is clearly seen in the bottom panels of \figu\ref{fig|GammaDependenceHalo} in which we plot the same models as in the upper panels but restricted in the narrow bins of stellar mass and effective radius, $12.05<\log \mstare/\msune<12.15$ and $1.2<\log R_{\rm e}/{\rm kpc}<1.3$. In this case the trend of \gamm\ with halo mass gets somewhat flatter, but still highly significant $p=-0.21\pm0.02$.

While \eq\ref{eq|gammaMhalo} has a relatively weak dependence on galaxy properties, it has a significant dependence on the choice of input NFW dark matter halo profile. The left, middle, and right panels of \figu\ref{fig|GammaDependenceHalo} show predictions for the same models assuming dark matter halo profiles uncontracted, with Adiabatic Expansion, and Adiabatic Contraction, respectively. It is clear that irrespective of the chosen stellar and/or dark matter mass profiles, and/or stellar mass-halo mass mapping, the predicted values of \gamm\ tend to decrease at broadly the same pace with increasing halo mass, with only a small offset in normalization of $\lesssim 15\%$. This behavior is mainly induced by structural non-homology, with the relative density distribution of dark matter becoming gradually more dominant in cluster-sized halos. Despite decreasing with increasing halo mass, the predicted values of \gamm\ still tend to be steeper than those actually measured in haloes above $\mhaloe\gtrsim 3\times10^{14}\, \msune$ by \citet{Newman13b}. Including adiabatic expansion in the models yields a stronger decrease of \gamm\ with effective radius, mirroring the flattening of the mass profile at $r\gtrsim R_{\rm e}$ (see \sect\ref{subsec|MassProfileDependenceSizeMass}), in apparent better agreement with the data. Still, we do not properly reproduce the total mass density profiles derived by \citet{Newman13b} for dominant galaxies in rich clusters.

\begin{figure*}
    \center{\includegraphics[width=17truecm]{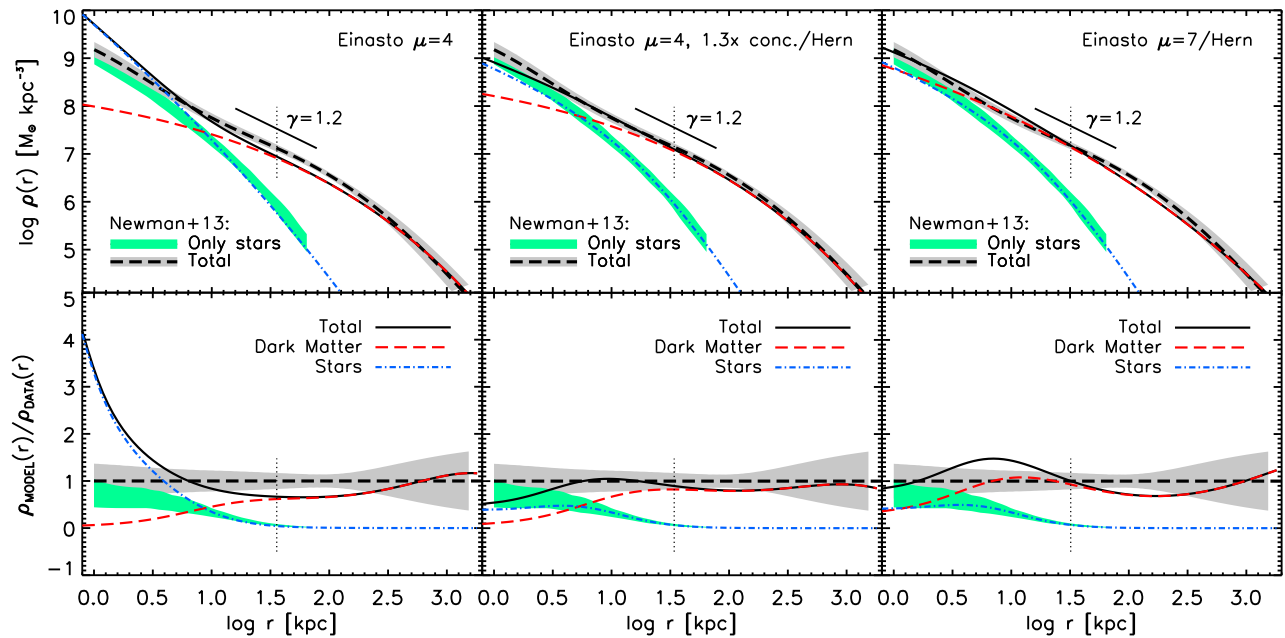}
     \caption{Same format as \figu\ref{fig|GammaInClusters} showing models characterized by a S\'{e}rsic+Einasto with $\mu=4$ (left),  a Hernquist+Einasto with $\mu=4$ and a mean concentration higher by a factor of $1.3$ (middle), and a Hernquist+Einasto with $\mu=7$ (right). The match to the data in this case does not require a prominent core nor a much higher median dark matter concentration.\label{fig|GammaInClustersConc}}}
\end{figure*}

\figu\ref{fig|GammaInClusters} reports a closer comparison between model predictions and the total mass density profile measured by \citet{Newman13b}. In all top panels of \figus\ref{fig|GammaInClusters} and \ref{fig|GammaInClustersConc}, the black long-dashed line and gray area mark, respectively, the \citet{Newman13b} mean total mass density profile and 1$\sigma$ dispersion, while the light green area roughly marks their $1\sigma$ dispersion in the BCG stellar profile. The bottom panels of \figus\ref{fig|GammaInClusters} and \ref{fig|GammaInClustersConc} plot instead the ratios between model predictions and data for each model. In all panels the predicted stellar and dark matter contributions to the total mass density profiles are reported by blue triple dot-dashed, and red long-dashed lines, respectively. The left panel in \figu\ref{fig|GammaInClusters} shows predictions for the reference stellar S\'{e}rsic and dark matter NFW profiles. There are two main issues. The reference model is much steeper than the data at small scales, severely overpredicting them at $r \lesssim 4$ kpc by up to a factor of $\sim 3-4$. At intermediate scales $10\lesssim r \lesssim 300$ kpc the model instead significantly underpredict the data at by a systematic factor of $\sim 2$.

Switching to a Hernquist profile in the stellar component naturally produces profiles at small scales much better aligned with the data (middle panel of \figu\ref{fig|GammaInClusters}). The resulting stellar profile is indeed in good agreement with the stellar profile inferred by \citet[][light green areas]{Newman13b}, predicting a density of $\rho_{\rm star} \sim 3\times 10^8\, \mstare/{\rm kpc^3}$ at $r \sim 2$ kpc, and $\rho_{\rm star} \sim 1.5\times 10^7\, \mstare/{\rm kpc^3}$ at $r \sim 10$ kpc. A Hernquist profile is a simple but clearly not a unique solution. Adding, for example, an appropriate ``core'' radius \citep[e.g.,][]{Trujillo04,Newman13b} to the stellar profile could yield similar improvements. A flatter core in the inner regions of very massive galaxies, especially at the center of clusters, has been indeed reported several times in the literature \citep[e.g.,][]{Kormendy09,Postman12,Huang13,Oldham16}. On the other hand, simply assigning a lower S\'{e}rsic index of, say, $n\lesssim 3$ to the most massive galaxies, although flattening the inner density profile (cfr. \figu\ref{app|Sersic3Dprofile}), would also decrease the predicted stellar mass density profiles in more external regions around a few to ten kiloparsecs, worsening the match to the galaxy stellar profiles measured by \citet{Newman13b}. Moreover, detailed light profile fits to massive galaxies have revealed their S\'{e}rsic indices to be as large as $n \gtrsim 5$ \citep[e.g.,][]{Huertas13b,Bernardi12}.

Despite the possible improvements at small scales, our reference models still significantly underpredict the mass density observationally inferred by \citet{Newman13b} at intermediate scales beyond $\gtrsim 10$ kpc, where the mass profile is expected to become fully dominated by the dark matter component. We find that additionally allowing for a boost of a factor of $2$ in the median halo concentration helps resolving the mismatch with the data, as shown in the right panel of \figu\ref{fig|GammaInClusters}. Increasing the concentration in fact decreases the core radius $r_s=r_{200c}/c_{200c}$ at fixed halo mass and virial radius $r_{200c}$, thus naturally increasing the mass density in the inner regions, with the end result of flattening the total mass density profile, especially around the effective radius (vertical, dotted lines in \figu\ref{fig|GammaInClusters}). However, we note that an increased concentration also tends to predict a larger mass density at $r \lesssim 10$ kpc. To avoid the latter effect, when adopting a higher concentration we must also allow for a core in the dark matter profile, for which we adopt \eq\ref{eq|coredNFW}. We checked that also adiabatic expansion could reduce the dark matter in the inner regions at comparable levels to those of a core.


Is there any observational justification for increasing concentrations over the predictions of N-body simulations? An increase in dark matter concentration at cluster scales is indeed possible \citep[e.g.,][and references therein]{Come07,Auger13}, though at present good evidence for real clusters to be on average more concentrated is still not convincing \citep[e.g.,][and references therein]{Ettori10,DuttonMaccio,Mene14,Merten14,Shan15,Umetsu15}.
\citet{Newman13b}, and more recently \citet{Amodeo16} at $z\gtrsim 0.4$, empirically constrained the median concentration at $\mhaloe \sim 10^{15}\, \msune$ from their samples to be $c\sim 5-6$, in broad agreement with our inferred values of $c\sim 2\times 3\div4 \sim 6\div8$,
which are consistent with the numerical results by \citet[][but see \citealt{MeneghettiRasia13}]{Prada12}. Evidence for larger concentrations for red, massive galaxies consistent with what presented in \figu\ref{fig|GammaInClusters}
also come from the kinematics of satellites, though mostly inferred from isolated galaxies \citep{WM13}.
\citet[][see also \citealt{Giocoli16}]{Giocoli14} discuss how selection effects
may bias the measured concentration$-$mass relation at cluster scales, yielding up to 30\% higher concentrations
than what predicted by N-body simulations \citep[see also][]{Auger13,Sereno15,Lieu17}.
Several studies have also highlighted the possibility of biases arising from halo triaxiality \citep[e.g.,][]{Foex14}.
An alignment of the major axis and the line of sight may cause an apparent higher concentration, that may be
evident in lensed groups and clusters. Although the \citet{Newman13b} cluster sample is relatively small,
the selection is well defined, and in fact \citet{Newman13b} find that only in one case there is substantial elongation along the line of sight (A383)
which could artificially boost the measured concentration.

\begin{figure*}
    \center{\includegraphics[width=13truecm]{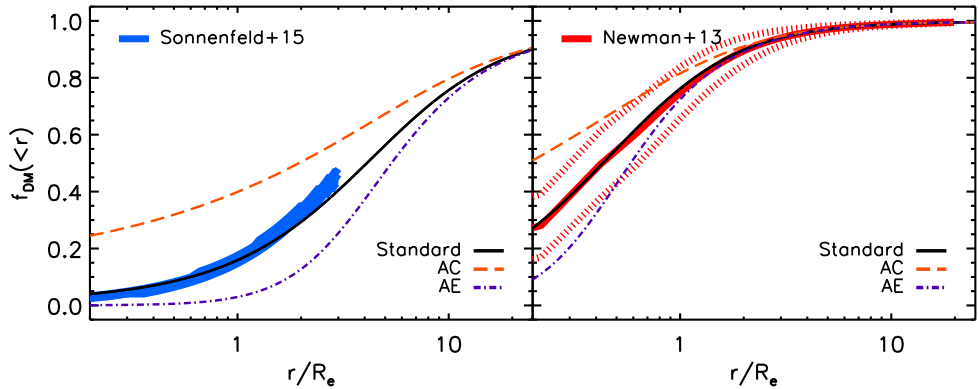}
    \caption{Ratio of dark matter mass and total (dark+stellar) mass as a function of scale normalized to the effective radius
    for galaxies in the range $11.3<\log \mstare/\msune<11.5$ and $6<R_{\rm e}/{\rm kpc}<7$ (left) and $\log \mstare/\msune>12$ (right). The black solid, orange long-dashed, and purple dot-dashed lines show predictions for the uncontracted NFW profile, with Adiabatic Contraction, and Adiabatic Expansion, respectively. In both panels show predictions for  standard S\'{e}rsic+NFW profiles. Overall, the data tend to disfavor any adiabatic contraction or expansion. \label{fig|fDM}}}
\end{figure*}

As numerical simulations and direct observations do not tend to
unanimously support dark matter concentrations in cluster-sized halos much higher than what predicted by numerical simulations, we have explored in \figu\ref{fig|GammaInClustersConc} the effects of switching from a NFW to an Einasto profile, which is characterized by an additional parameter $\mu$ (\eq\ref{eq|Einasto}), and it has been shown to be an accurate fit to massive dark matter halos by a number of groups \citep[e.g.,][]{Navarro04,Stadel09,Navarro10,Ludlow13,DuttonMaccio,Klypin16}. The left panel of \figu\ref{fig|GammaInClustersConc} shows the results of a standard S\'{e}rsic+unperturbed dark matter Einasto profile with $\mu=4$ (or $\alpha=1/\mu=0.25$), which is the average value extracted from numerical N-body simulations of very massive clusters \citep[e.g.,][]{Navarro04,Gao08,DuttonMaccio}.  This basic model tends to present broadly the same shortcomings as the standard NFW-based model presented in the left panel of \figu\ref{fig|GammaInClusters}. However, the match is substantially improved at \emph{all} scales when simply allowing for a Hernquist stellar profile and a median concentration just 30\% higher than the reference numerical value (middle panel). A higher value of $\mu=7$ (right panel) further increases the dark matter density at intermediate scales without the need to invoke any increase in median concentration, but the match to the data in this case is poorer.

In summary, the present available lensing data for galaxies at the center of cluster-scale halos \citep{Newman13b}, tend to favor models with profiles in their inner stellar component at $r\lesssim 3-5$ kpc, flatter than what predicted by extrapolations of their outer S\'{e}rsic profiles. The same data also tend to align better with models characterized by a NFW profile with a core and a median concentration in the dark matter component twice the one predicted by dark matter-only N-body simulations, or by an Einasto profile with $\mu\sim 4$ and a dark matter concentration just $\lesssim 30\%$ higher than the simulation results. The latter results are in nice agreement with \citet{Auger13} who also found 26 group- and cluster-scale strong gravitational lenses to be only $\sim 26\%$ more overconcentrated than similar-mass halos from dark matter simulations. We stress that the relatively small comparison observational sample prevent us to generalize our findings to all galaxy clusters. The idea that Einasto models may be more accurate representations of the mass density profiles of especially the most massive halos is in line with a number of analytic and numerical studies \citep[e.g.,][]{Zhao03,Gao08,LapiCav11,CenSersic,Nipoti15,Angulo16,Klypin16,LudlowAngulo}. The common theme in these works is the connection between the final shape of the system, and thus of its parameter $\mu$, with its mass accretion history, in a way that $\alpha=1/\mu$ and ``curvature'' tend to increase with fast growth, as it might be the case, on average, for the inner regions of very massive halos.

\subsection{Dark Matter Fractions}
\label{subsec|DMfractions}

Up to this point, we have mainly discussed the outputs of semi-empirical models
with respect to \gamm, but clearly additional observables can and must be used to place tighter constraints on galaxy evolution.
In particular, scaling relations with velocity dispersions \citep[e.g.,][]{Dutton13}, evolution of the fundamental plane \citep[e.g.,][]{Treu01,ShankarBernardi09,Shankar13,Oldham17}, evolution of the virial relations \citep[e.g.,][]{Peralta14}, building up of age/metallicity gradients \citep[e.g.,][]{Montes14}, are all relevant issues to test models.

While we aim to explore at least some of these additional issues in separate work,
we here begin by comparing in \figu\ref{fig|fDM} our predicted 3D ratio of dark to total (dark plus stellar) mass as a function of scale (radius normalized to the effective radius) for all galaxies with $11.3<\log \mstare/\msune<11.5$ and $6<R_{\rm e}/{\rm kpc}<7$ (left panel) and $\log \mstare/\msune>12$ (right panel), compared with the lensing-based dark matter fractions derived from the \citet[][thick blue lines, left panel]{Sonne15} sample and \citet[][thick solid and dotted red lines, right panel]{Newman13b} galaxy samples.
Individual dark matter fractions in the \citet{Newman13b} cluster central galaxies are calculated by integrating the density profiles of their \figu3. We then take the mean of the dark matter fractions at each radius to produce the solid and dotted thick red lines reported in the right panel of \figu\ref{fig|fDM}. The black solid, orange long-dashed, and purple dot-dashed lines show predictions for the uncontracted NFW profile, with adiabatic contraction, and adiabatic expansion, respectively.
Both panels show predictions for our standard stellar S\'{e}rsic and dark matter NFW profiles. It is apparent that at cluster scales (right panel) the dark matter fractions rise relatively more rapidly than for lower mass galaxies in the lower mass halos (left panel), irrespective of the details in the stellar and/or dark matter profiles. This trend is mostly induced by the fact that more massive galaxies, characterized by progressively higher effective radii, naturally will include more dark matter mass in their central regions.

Our semi-empirical models, within the uncertainties, align with the observationally-based dark matter fractions, and we verified this broadly continues to hold true even with flat inner stellar profiles, and/or with cores in the dark matter component. In this respect, dark matter fractions appear to be less constraining for models, though adiabatic contraction or expansion continue to be disfavored by the data irrespective of the exact inputs in the semi-empirical models, in line with \citet{Oguri14}. Our predicted dark matter fractions are in full agreement with the $f_{\rm dm}(R_{\rm e})\sim 20-30\%$ values inferred by \citet[][see also \citealt{WM13}]{Cappellari13b} for $\mstare\sim 2-3\times10^{11}\,\msune$, or on larger scales with \citet{Alabi17}, who infer in the SLUGGS survey $f_{\rm dm}(<5R_{\rm e})\gtrsim 60\%$ from globular cluster kinematics data. At larger masses, our models predict $f_{\rm dm}(R_{\rm e})\sim 60-70\%$ for $\mstare\gtrsim 10^{12}\, \msune$, consistent with the extrapolation of the \citet{Cappellari13b} analytic fit, $f_{\rm dm}(R_{\rm e})\sim 1.3+0.24(\log \mstare-10.6)^2$, to the most massive galaxies in our mocks.

For completeness, to visualize the full dependence of the mass density profiles on the relative contributions between stellar and dark matter components, in \figu\ref{fig|gammavsfDM} we plot \gamm\ as a function of dark matter fractions $f_{\rm dm}(<R_{\rm e})$ within one effective radius for galaxies with $\mstare \gtrsim 3 \times 10^{11}\, \msune$ as predicted by our reference semi-empirical model with uncontracted NFW (solid and dotted lines), with adiabatic contraction (orange, dashed line), and with adiabatic expansion (purple, dot-dashed line). All models predict a steady drop of \gamm\ with increasing dark matter fraction, as expected given that, as discussed with respect to \figu\ref{fig|fDM}, larger $f_{\rm dm}(<R_{\rm e})$ tend, on average, to be linked to larger effective radii with more prominent dark matter contributions, and thus flatter \gamm, also in line with \figu\ref{fig|GammaReCluster}. However, models with adiabatic expansion tend to predict a substantially faster drop with $f_{\rm dm}(<R_{\rm e})$ than the other two models. Our uncontracted NFW predictions nicely match those from
the dynamical modeling on ATLAS3D and SPIDER galaxies by \citet[][empty squares with error bars]{Tortora14a} who also assume an underlying NFW profile. On the other hand, while model predictions agree with the dark matter fractions extracted from \citet[][blue squares]{Sonne15}, they are highly inconsistent with the dark matter fractions derived from \citet[][red circle]{Newman13b} for galaxies in clusters. As already emphasized for \figu\ref{fig|GammaReCluster}, galaxies characterized by similar stellar mass, effective radius
and even dark matter fractions tend to significantly differ in their total mass density slopes, an additional sign that host halo mass and its (outer) structure is systematically different.

\section{Discussion}
\label{sec|Discu}

\subsection{Constraints from the dependence of \gamm\ on size and stellar mass}

We have carried out a detailed comparison of a set of semi-empirical models with a diverse set of observational data.
The models are based on assigning galaxies, with measured stellar mass, effective radius and S\'{e}rsic index, to dark matter halos via abundance matching techniques. By design, our approach is based on a minimal number of parameters and underlying assumptions, mainly related to the underlying dark matter mass distribution, which is relatively well constrained by high-resolution N-body simulations \citep{Prada12,DuttonMaccio,Bene14}, and the IMF, which we usually set equal to a Salpeter or even variable IMF, having shown in \figu\ref{fig|GammaPosteriorProbIMF} that a Chabrier IMF is ruled out by the data in massive galaxies, in line with a number of independent studies based on surface gravity-sensitive absorption lines in stellar population synthesis models \citep[e.g.,][]{ConroyDokkumIMF,LaBarbera13,Spiniello14}, and/or stellar kinematics also coupled to strong lensing measurements \citep[e.g.,][]{Auger10,Treu10,Spiniello11,Thomas11,Barnabe13,Cappellari13b,Posa15}.

\begin{figure}
    \center{\includegraphics[width=8.5truecm]{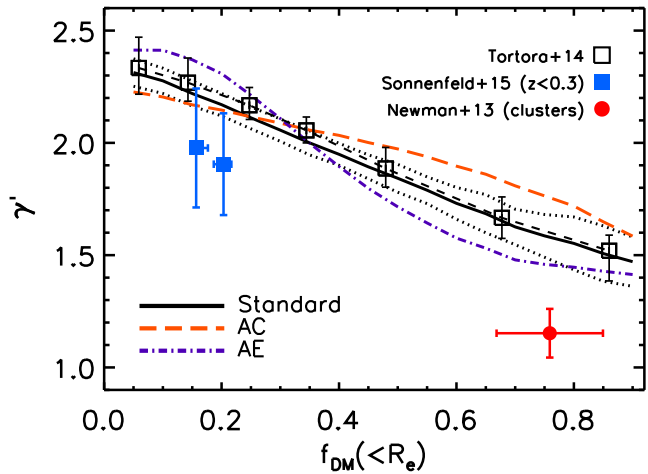}
    \caption{Dependence of \gamm\ on dark matter fraction within one effective radius for galaxies with $\mstare \gtrsim 3 \times 10^{11}\, \msune$ as predicted by our reference S\'{e}rsic+NFW semi-empirical model with uncontracted NFW (solid and dotted lines), with adiabatic contraction (orange dashed line), and with adiabatic expansion (purple dot-dashed line). Empty squares are the results of the dynamical modeling on ATLAS3D and SPIDER by \citet{Tortora14a}, blue squares from \citet{Sonne15}, and red circle from \citet{Newman13b} for galaxies in clusters. It is clear that the latter subsample of galaxies is significantly off with respect to galaxies of similar dark matter fraction, indicating that galaxies in rich clusters are systematically different. \label{fig|gammavsfDM}}}
\end{figure}

Our semi-empirical models thus represent ideal tools to probe the consistency among independent data sets,
to reliably test the predictions of dark matter numerical simulations, to pin down the role of adiabatic contraction/expansion, and environment in shaping the total mass density profiles of massive galaxies. The semi-empirical models successfully reproduce the total mass density profile of massive galaxies with $\mstare \gtrsim 3\times 10^{11}\, \msune$ (Salpeter IMF), along with its dependence on stellar mass and effective radius, at least in non-cluster environments. We find that even moderate modifications to the halo profile in terms of adiabatic contraction/expansion are disfavored at high significance. In the next steps (Shankar et al. in prep.) we will mainly focus on the evolution of \gamm\ as a function of redshift, as well as the additional constraints that can be gained by a comprehensive Jeans-based calculation of velocity dispersions \citep[e.g.,][]{Chae14}.

Our results on the dependence of \gamm\ on galaxy properties are in line with several previous works. We discuss some of these here below.

\citet{Xu17} have recently compared the properties of simulated early-type galaxies in the Illustris hydrodynamic cosmological simulation with the same SLACS data as in this paper. Their simulations broadly align with observations, suggesting a flattening of the total mass density profile with increasing effective radius, though generally shallower than what the data suggest.

\citet{Poci17} have also recently confirmed from ATLAS3D that the slopes of the total mass density profiles are approximately isothermal, and depend on various galactic properties, most significantly on mass surface density, while the correlation with velocity dispersion becomes negligible above $\log \sigma/{\rm km\, s^{-1}}\gtrsim 2.1$.

\citet{Tortora14a} have explored the density slope
dependence in early-type galaxies over quite a broad range of stellar mass finding clear signs of
non-universality. They suggested that the relation with effective radius is the most robust to changes in the underlying modeling.
They find, in fact, that the largest galaxies have a density slope close to isothermal, while the smallest ones are characterized by a steeper
profile, in line with our findings.

\citet{DuttonTreu} have also addressed the question of dependencies of \gamm\ on galactic properties, though
adopting only one stellar mass-halo mass mapping. In agreement with our work, they disfavor models with significant adiabatic contraction or expansion, and also predict an overall non-universal \gamm\ dependence on galactic properties.

\citet{Auger10IMF} compared data on 53 massive early-type lensed galaxies with a variety of semi-empirical models with varying baryon efficiency, adiabatic contraction, and stellar initial mass function. They highly disfavored light IMFs, such as a Chabrier of Kroupa, as well as strong adiabatic contraction, in nice agreement with the results presented in this work. Their models also suggest that only mass-dependent IMFs could be consistent with no adiabatic contraction, a conclusion not necessarily supported by our findings (see, e.g., \figu\ref{fig|App:GammaPosteriorProbIMFvar}).

\citet{JiangKoch} also carried out semi-empirical galaxy-halo modeling treating the stellar mass-to-light ratio as a free parameter.
They analyzed surveys of gravitational lenses with velocity dispersion measurements assuming combinations of Hernquist and NFW profiles.
Their total mass density slopes are peaked around the isothermal values with non negligible variations around the mean. They also found that models without adiabatic contraction favor a stellar fraction of around 3\%, in agreement with weak lensing estimates \citep[e.g.,][]{Mandelbaum06} and the most recent abundance matching results by \citet{Shankar14b} considered in this work (\figu\ref{fig|MstarMhalo}).

\citet{Dutton11} modeled the scaling relations of early and late-type galaxies, finding that massive galaxies characterized by a Salpeter IMF are not consistent with halo contraction. \citet{Napolitano10} also emphasized that that the lower central dark matter fractions inferred in the oldest early-type galaxies are inconsistent with adiabatic contraction.

By analyzing a suite of simulated spheroids formed out of binary mergers, \citet{Remus13} concluded that the exact
value of the density slope depends on the amount of gas involved in the merger, with more dissipative gas-rich mergers
producing steeper profiles. This would naturally imply some non-homology,
as more dissipative mergers would produce more compact galaxies at fixed stellar mass with steeper slopes, in line with what observed.
Dissipative processes may however not be the only cause for the dependence of \gamm\ on galactic properties.
Indeed, as discussed in \sect\ref{subsec|MassProfileDependenceSizeMass},
in basic $\Lambda$CDM models with uncontracted halos, the trend of \gamm\ with effective radius is simply controlled by the rate of progressive change from the steeper stellar component to the flatter dark matter one.

\citet[][see also \citealt{Sonne14merge,Dubois13}]{Remus16} also recently emphasized that the decreasing importance of dissipative processes towards lower redshifts and for more massive systems
is mostly responsible in shaping the strong flattening of \gamm\ with increasing dark matter fractions $f_{\rm dm}(<R_{\rm e})$ within one effective radius as measured by \citet{Tortora14a}. As detailed in \figu\ref{fig|gammavsfDM}, galaxies in clusters do however show significantly smaller \gamm\ distributions at fixed $f_{\rm dm}(<R_{\rm e})$, supporting a scenario in which also the larger scale environment plays a significant role in defining the total mass density profiles of early-type, massive galaxies.

\subsection{Constraints from the dependence of \gamm\ on halo mass}

We showed that the mass-weighted total mass density slope at the effective radius \gamm\ is a strong decreasing function of the host halo mass, ranging from values close to, or higher than, the canonical isothermal value of \gamm\ $\sim 2$ in the field/low-mass groups, to values closer to \gamm\ $\sim 1.5$ at the cluster scale. This behavior is mainly induced by the increasing
contribution of the dark matter component around the effective radius,
thus inducing \gamm\ to progressively flatten out (cfr. \figu\ref{fig|LocalRhoMr}).

A closer comparison to the data interestingly reveals that for very massive galaxies with $\log \mstare/\msune \gtrsim 11.9$ in cluster-scale environments (cfr. \figu\ref{fig|GammaInClusters}), the stellar profile at small scales $r \lesssim 3-5$ kpc should be significantly flatter than what predicted
by an extrapolation of a S\'{e}rsic profile. This is in line with several independent lines of evidence \citep[][]{Postman12,Kormendy09}, as mentioned in \sect\ref{subsec|MassProfileDependenceEnvironment}.

In general, several physical processes are thought to play some role in shaping the density profiles of massive galaxies, especially in overdense regions.
\citet{Laporte14}, for example, have presented a series of collisionless $N$-body resimulations
to follow the growth of central cluster galaxies and their dark matter halos since redshift $z\lesssim 2$.
In line with what discussed in this work, \citet{Laporte14} report a steepening at small scales in the stellar component not mirrored in the data. They propose that the action of repeated super-massive black hole mergers could create
enough large cores of a few kiloparsecs to alleviate the tension between models and observations \citep[see also, e.g.,][]{Merritt06}.

However, other in-situ mechanisms may be able to induce a permanent flattening of the central matter distributions,
such as the action of efficient feedback \citep[e.g.,][]{Peirani08,Martizzi13} from central active galactic nuclei (AGN).
On the other other hand, an ``in-situ'' process such as AGN feedback, a phenomenon more and more supported by direct and indirect observations \citep[e.g.,][and references therein]{Shankar16BH,Shankar17BH,Barausse17}, should produce a stellar core in all massive galaxies, irrespective of their environment.

\citet{Laporte14} also discuss that multiple dissipationless mergers can effectively flatten out the inner distributions,
with the dark matter component producing a shallower central cusp, and gravitational heating redistributing material
in the outer regions \citep[see also, e.g.,][]{Elzant04,Nipoti04,ToniniLapi,Lapi09b,Elzant16}.
This type of evolution is in line with our proposed modifications presented in \figu\ref{fig|GammaInClusters}
of a cored NFW profile to faithfully align model predictions to observations in the range $3<R/\kpce<20$.

Finally, we note that other non-baryonic processes mainly linked to the dark sector, such as self-interacting dark matter, may contribute to the diversity in total mass density profiles inferred in this work \citep[e.g.,][]{Kapli16,DiCintio17}.

\citet{Chae14} fitted two-component models to local SDSS nearly spherical galaxies. Adopting a velocity dispersion-dependent variable IMF, they found a mean slope for the total mass density profiles of $\langle\gamma_e\rangle=2.15\pm 0.04$, computed in the radial range $0.1\ree<r<\ree$, in full agreement with our findings. Chae et al. also went forward in testing the dependence on galactic properties including effective radius and halo mass. Their least-square fit total mass density slope between $0.1\ree$ and \re\ also decreases with effective radius, though more weakly
than the one reported in \figu\ref{fig|GammaReCluster}. \citet{Chae14} also infer a dependence of the total mass density slope on host halo mass weaker than the one measured here, though they also show that a much stronger dependence is naturally obtained when the slope of the mass density profile is fitted on scales larger than their effective radius.

\section{Conclusions}
\label{sec|Conclu}

In this work we have set up a semi-empirical approach to create large catalogs of
mock galaxies and host dark matter halos at $z=0.1$.
To this purpose, we have selected a large sample of galaxies from SDSS with a probability $P(E)>0.85$ of being ellipticals,
and with measured stellar masses, effective radii, and S\'{e}rsic indices. We then assigned host dark matter halos to each galaxy in the sample
making use of the most recent stellar mass-halo mass relations \citep{Shankar14b}, as well as exploring the impact of different mappings. We finally computed for each
galaxy the total mass density profile and compared with a variety of data sets from strong and weak lensing as cataloged and analyzed by \citet{Sonne13},
\citet{Newman15}, and \citet{Newman13b}, and also, where appropriate, with dynamical measurements from \citet{Cappellari15}.
Our main results, for galaxies with mass above $\mstare \gtrsim 2-3 \times 10^{11}\, \msune$ (Salpeter IMF), can be summarized as follows:
\begin{itemize}
  \item In line with observational evidence, the semi-empirical models naturally predict a non-universal mass weighted total density profile at the effective radius \gamm, independently of the exact input assumptions. In the specific,
      the models confirm the observational evidence that \gamm\ decreases for increasing galaxy size at fixed stellar mass, and for lower mass galaxies at fixed effective radius.
  \item The strong lensing data disfavor at $\gtrsim 3 \, \sigma$ a Chabrier IMF and are consistent with a Salpeter or variable IMF.
  \item Data also disfavor at $\gtrsim 2-3 \, \sigma$ departures from a S\'{e}rsic stellar profile and an uncontracted NFW dark matter profile in the inner regions.
  \item Our modeling predicts the total density profile to be roughly isothermal or steeper (\gamm$\gtrsim 2$) in low-mass halos of $\mhaloe \lesssim 10^{13}\, \msune$, but shallower (\gamm$\sim 1.5$) in cluster environments, with $\mhaloe \gtrsim 3\times 10^{14}\, \msune$. This is mainly a manifestation of structural non-homology, with the relative density distribution of stars and dark matter varying systematically from isolated galaxies to central galaxies in clusters. Despite this substantial decrease with increasing halo mass, the predicted \gamm\ are still significantly steeper than the total mass density slopes currently measured in BCGs, which suggest values closer to \gamm\ $\sim 1.1$.
  \item Adequately reproducing the full total mass density profiles of massive galaxies at the center of massive halos with $\mhaloe \gtrsim 3\times10^{14}\, \msune$ requires a stellar mass profile flatter than S\'{e}rsic within $r\lesssim 3-5$ kpc. An improved match at intermediate scale $10\lesssim r \lesssim 300$ kpc can instead be obtained by assuming: 1) either a cored NFW dark matter profile, coupled to a median halo concentration a factor of $\sim 2$ higher than the one predicted by N-body numerical simulations; 2) or an Einasto profile with $\mu \sim 4$ ($\alpha \sim 0.25$), coupled to a median halo concentration only a factor of $\lesssim 1.3$ higher than numerical predictions.
\end{itemize}

\section*{Acknowledgments}
F.S. acknowledges valuable discussions with A. Newman, C. Kochanek, A. Dutton, and S. Dye.

R.G. acknowledges support from the Centre National des Etudes Spatiales (CNES).
T.T. acknowledges support from the Packard Foundation through a Packard Research Fellowship.
We thank two anonymous referees for several useful comments and suggestions that greatly improved the quality and presentation of our results.
K.-H. C. was supported by Basic Science Research Program through the National Research Foundation of Korea  (NRF)  funded  by  the  Ministry  of   Education   (NRF-2016R1D1A1B03935804).

\appendix

\section{The concentration-mass relation}
\label{app|Conc}

\begin{figure}
    \center{\includegraphics[width=8.5truecm]{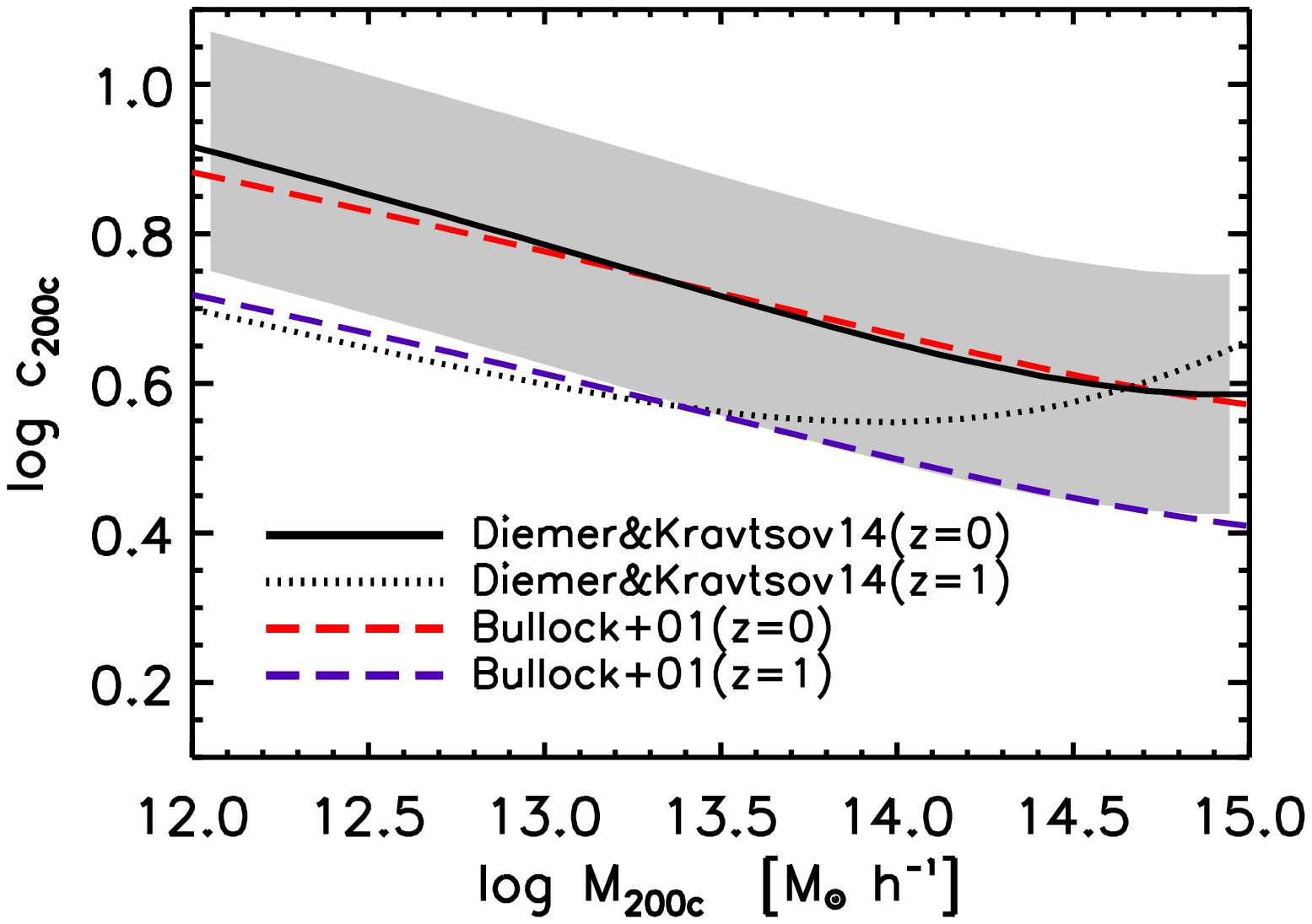}
    \caption{Concentration versus halo mass for our cosmology computed from the \citet{Bene14} model (solid line) along with its $1~\sigma$ uncertainty (gray region),
    also plotted at $z=1$ (dotted line). Also shown for comparison the results of the \citet{Bullock01} concentration model as revised by
    \citet{Maccio08} and \citet{DuttonTreu} (red and purple long-dashed lines) for the same cosmology, using $K_{200c}=4$ (see text for details).\label{fig|App:Concentration}}}
\end{figure}

The reference concentration-halo mass relation we always use in this paper has been derived by \citet{Bene14}.
Their results are shown in \figu\ref{fig|App:Concentration}, for our chosen cosmology (\sect\ref{sec|intro}), at $z=0$ (solid line), with the gray area marking the $1\sigma$ uncertainty region of 0.16 dex, and at $z=1$ (dotted line).

We compare the results by \citet{Bene14} with the analytic model put forward
by \citet{Bullock01} and further revised by \citet{Maccio08}.
The latter first look for the redshift of collapse $z_c$ of a given halo as the redshift at which the characteristic mass is equal to a fraction $F=0.01$ of the halo mass at the observation redshift $z$.
It then computes the concentration for halos defined as 200 times the critical density as
\begin{equation}
c_{200c}=K_{200c}\left[\frac{H(z_c)}{H(z)}\right]^{2/3}
\end{equation}
where $H^2(z)=H_0^2 [\Omega_{\Lambda}+\Omega_m(1+z)^3]$, and
we set $K_{200c}=4$ in line with the numerical results by \citet{DuttonMaccio}.

The \citet{Bullock01} and \citet{Maccio08} analytic model (red and purple long-dashed lines) is in very good agreement with the \citet{Bene14} results, but mostly at low redshifts, while it departs from it at increasing redshifts (dotted line; see also \citealt{Klypin16}).

\newpage

\section{The S\'{e}rsic 3D profile}
\label{app|Sersic3Dprofile}

Given a stellar mass, a projected effective radius $R_{\rm 2D}$, and a S\'{e}rsic index $n$ assigned to a SDSS galaxy from photometric image analysis,
we work out the full 3D S\'{e}rsic profile following, in the specific, the analytic fits by \citet[][]{Lima99},
although other groups have provided useful approximations \citep[e.g.,][]{Prugniel97,Mamon05}.
We first convert the input, empirical 2D radius to a 3D one using their approximation
 \begin{equation}
 r_H=R_{\rm 2D}(1.356-0.0293m+0.0023m^2)\, ,
 \label{app|R2DR3d}
 \end{equation}
with $m=1/n$.
The 3D mass density profile at any radius $r$ reads then as
 \begin{equation}
 f(x)=f_0 x^{-p} \exp(-x^m)\,\, ,
 \end{equation}
where
 \begin{equation}
 p = 1.0-0.6097m+0.05463m^2
 \end{equation}
and $x=r/a$, $\ln(a)=\ln(R_{\rm 2D})-k$, and $k=[0.6950-\ln(m)]/m-0.1789$.
The normalization $f_0$ is then fixed by combining Eqs.~9 and 18 in \citet{Lima99} as
 \begin{equation}
 f_0=\frac{\mstare}{4\pi a^3}\frac{m}{\Gamma[(3-p)/m]}\,\, .
 \label{eq|RhorSersic}
 \end{equation}
The cumulative mass $M(<r)$ within a given radius $r$ can be analytically expressed as
\begin{equation}
M(<r)=0.5\mstare\frac{\gamma[(3-p)/m,x^m]}{\gamma[(3-p)/m,(r_H/a)^m)]} \,
\label{eq|MrSersic}
\end{equation}
with $\gamma$ here indicating the (lower) incomplete gamma function.

\begin{figure}
    \center{\includegraphics[width=8.5truecm]{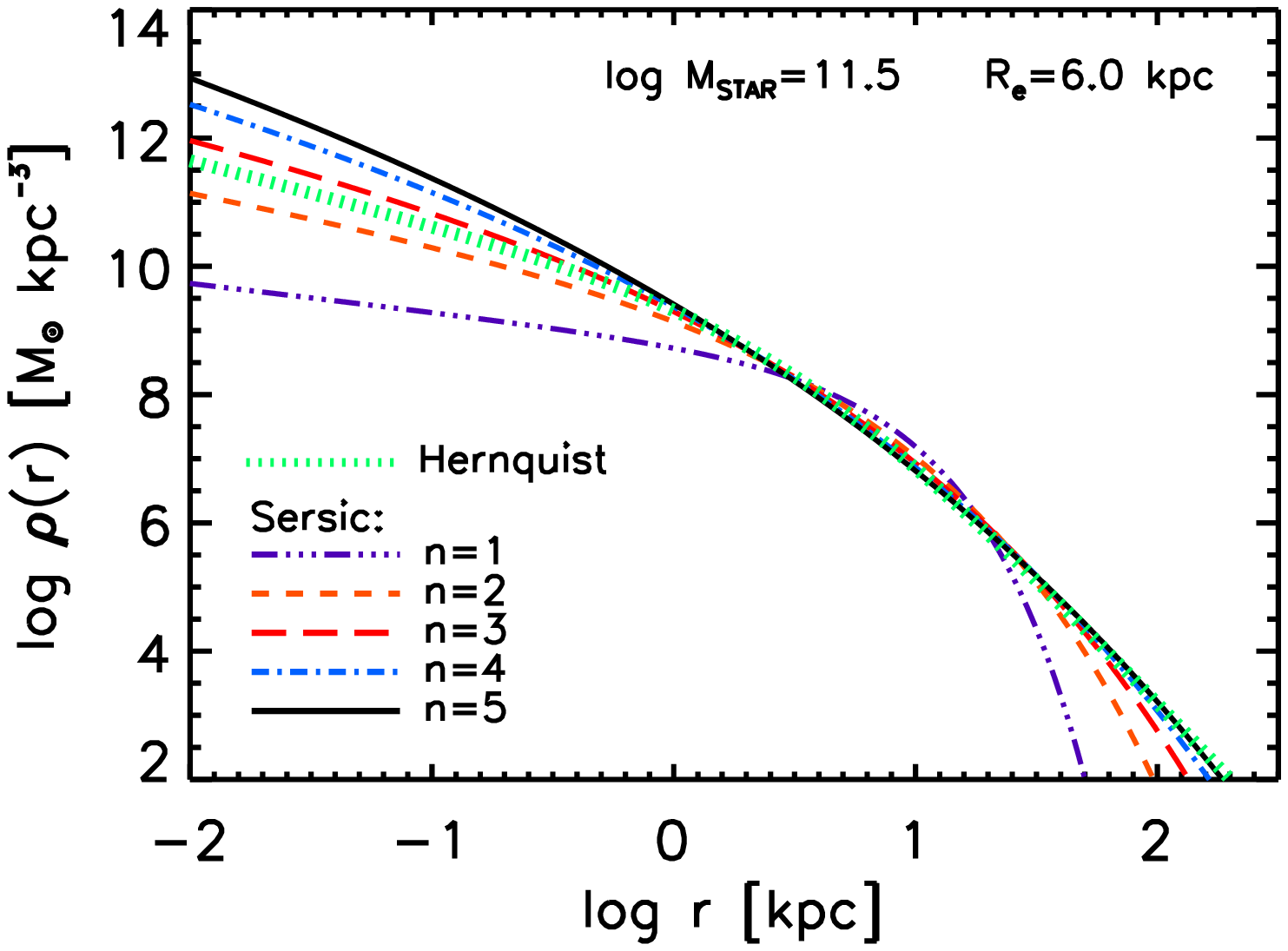}
    \caption{Schematic sketch of the change in stellar mass profile with decreasing S\'{e}rsic index $n$, as labeled, for a galaxy at fixed stellar mass and effective radius. The profile
    tends to flatten out with decreasing $n$ inducing a proportional flattening in \gamm. For comparison also shown the predicted Hernquist profile.\label{fig|App:SersicProfileExample}}}
\end{figure}

An example of the S\'{e}rsic profile for different values of the S\'{e}rsic index $n$ at fixed stellar mass and effective radius is given in \figu\ref{fig|App:SersicProfileExample}, where we also compare it to the Hernquist profile.

\section{Alternative models}
\label{app|OtherModels}

\begin{figure*}
    \center{\includegraphics[width=15truecm]{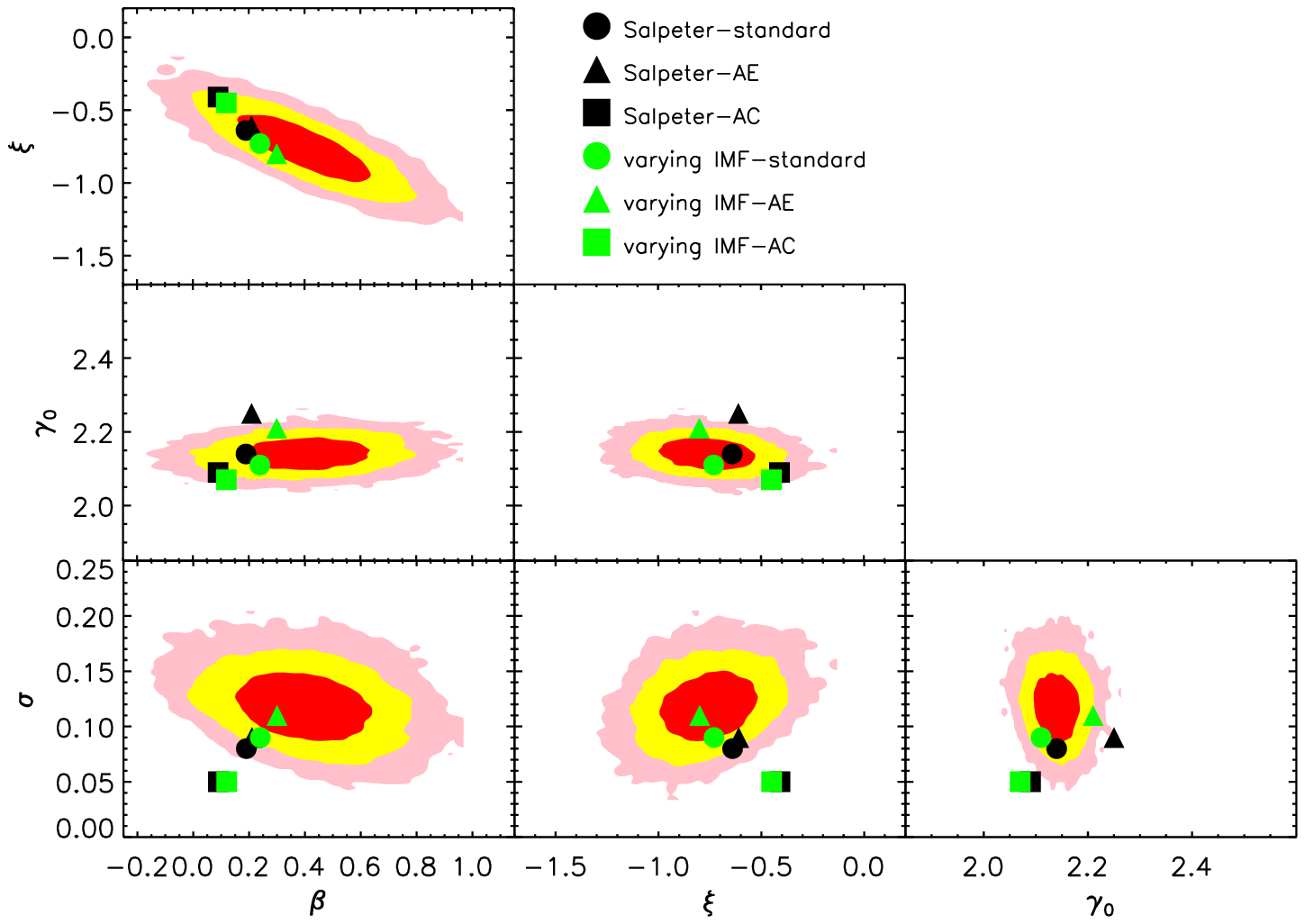}
    \caption{Same format as \figu\ref{fig|GammaPosteriorProbIMF} with the Salpeter IMF model compared with the variable IMF one.
    Both models perform .  \label{fig|App:GammaPosteriorProbIMFvar}}}
\end{figure*}

\begin{figure*}
    \center{\includegraphics[width=15truecm]{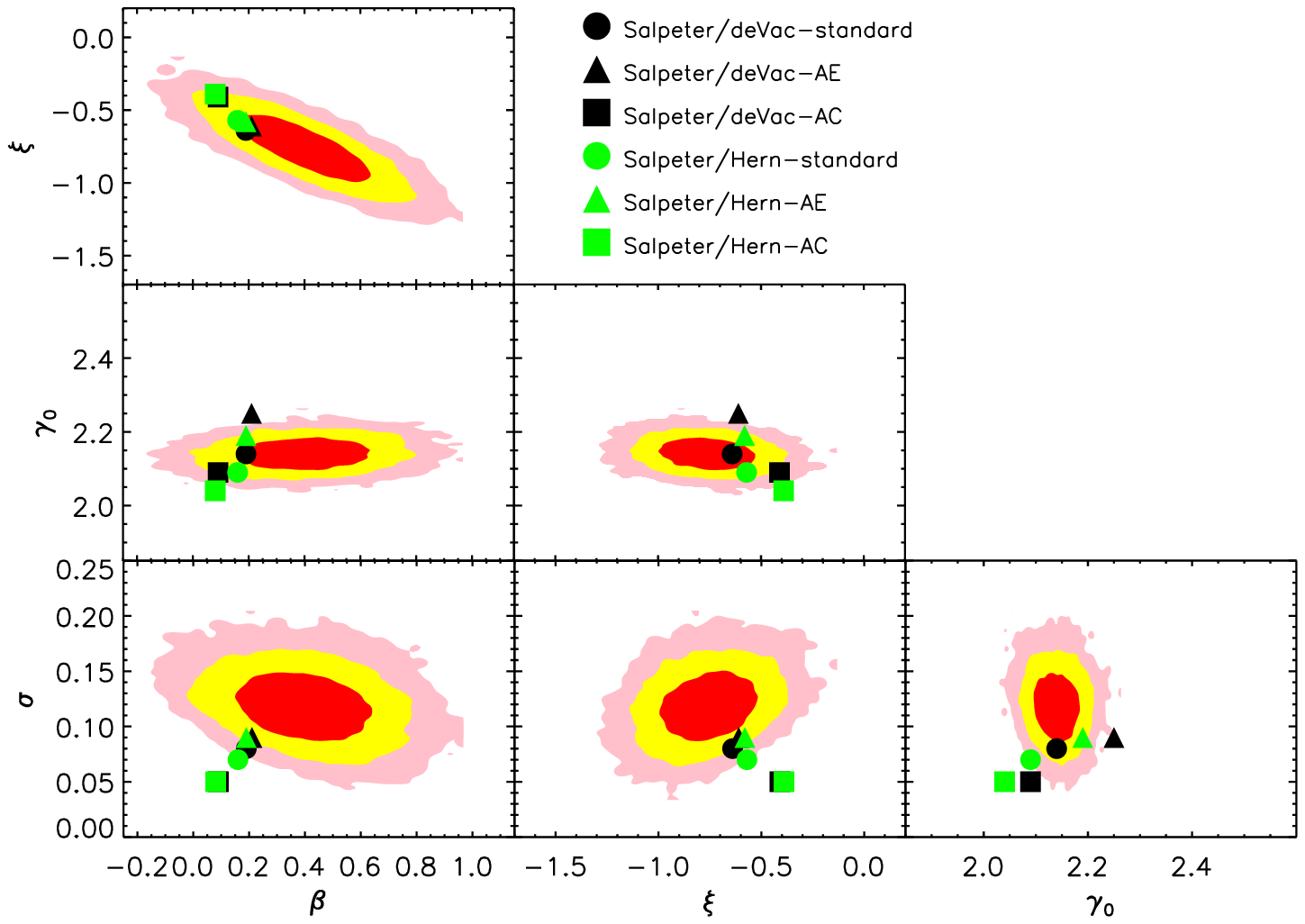}
    \caption{Same format as \figu\ref{fig|GammaPosteriorProbIMF} with the standard \devac\ model compared with a Hernquist stellar profile model.
    The latter model is less favored by the data, in line with what inferred from kinematic data \citep[][see \figu\ref{fig|LocalRhoMr}]{Cappellari15}. \label{fig|App:GammaPosteriorProbHern}}}
\end{figure*}

\begin{figure*}
    \center{\includegraphics[width=15truecm]{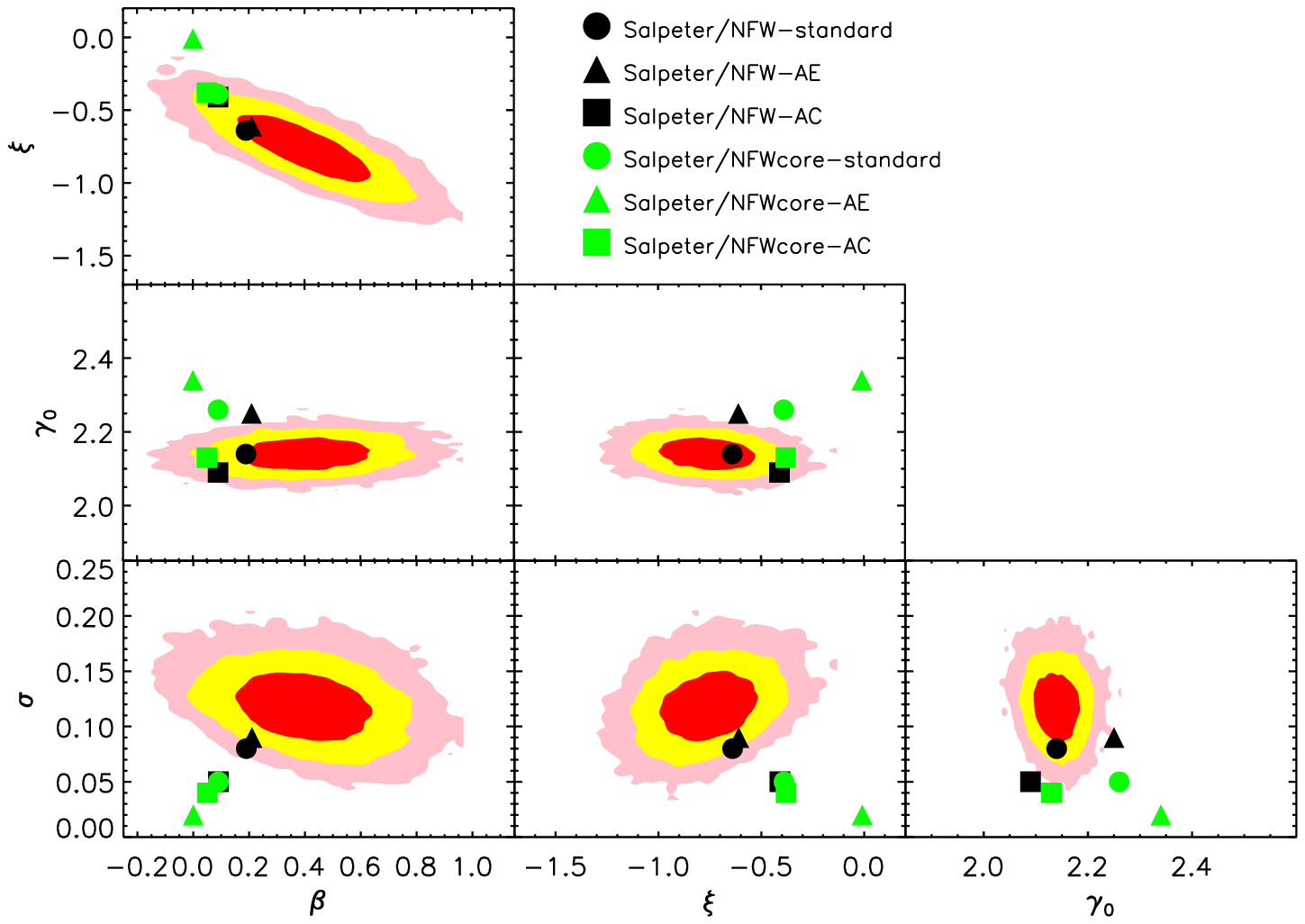}
    \caption{Same format as \figu\ref{fig|GammaPosteriorProbIMF} with the standard NFW model compared with a cored NFW model as in \citet{Newman13b}.
    The latter model is disfavored, at least at the stellar and halo mass scales probed by \citet{Sonne13}.
    \label{fig|App:GammaPosteriorProbCore}}}
\end{figure*}

\begin{figure*}
    \center{\includegraphics[width=15truecm]{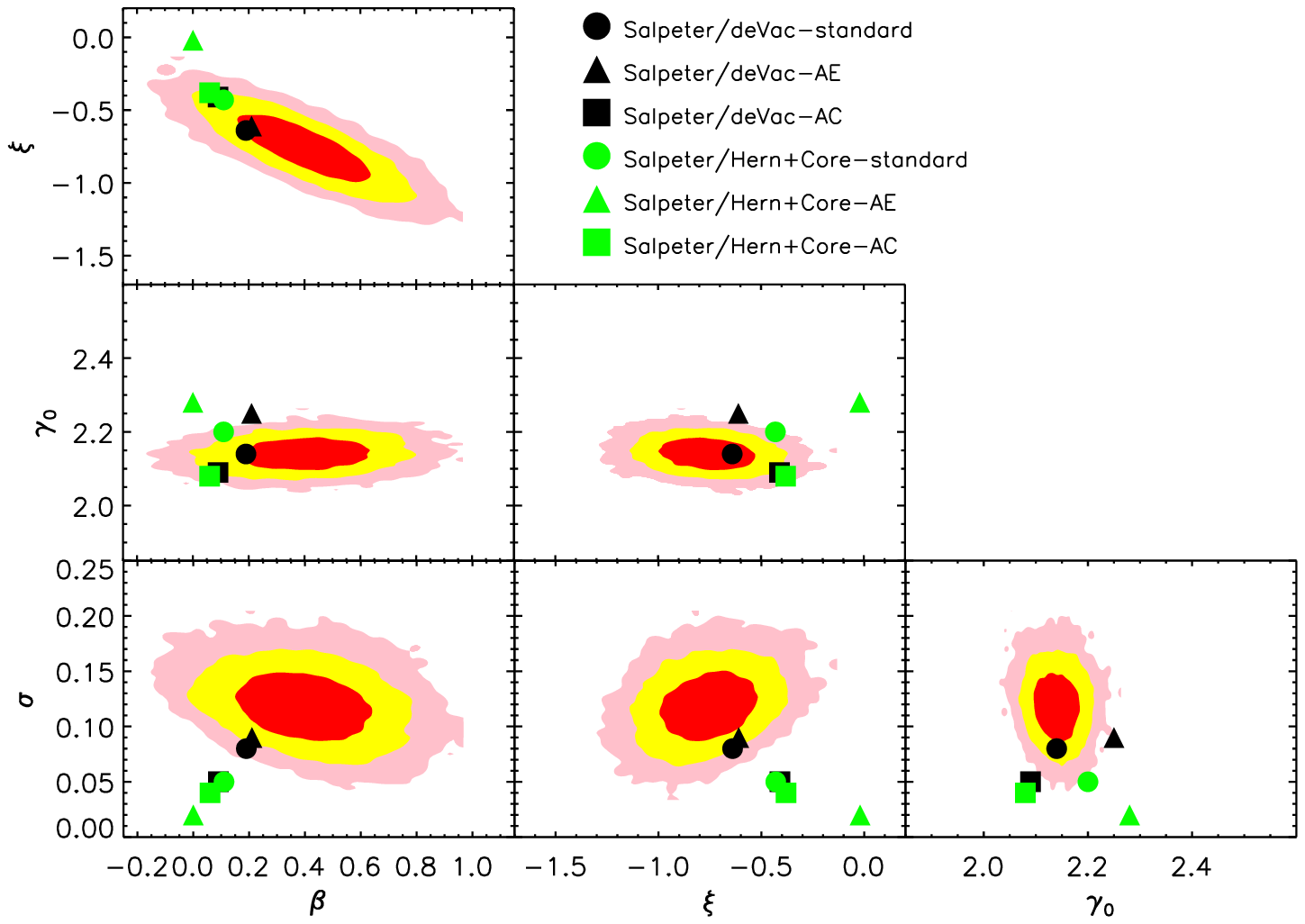}
    \caption{Same format as \figu\ref{fig|GammaPosteriorProbIMF} with the \devac+NFW standard model compared with a stellar Hernquist+cored NFW model.
    The latter model is disfavored in intermediate-mass galaxies with respect to a standard model with steeper stellar and dark matter profiles.  \label{fig|App:GammaPosteriorProbCoreHern}}}
\end{figure*}

\begin{figure*}
    \center{\includegraphics[width=15truecm]{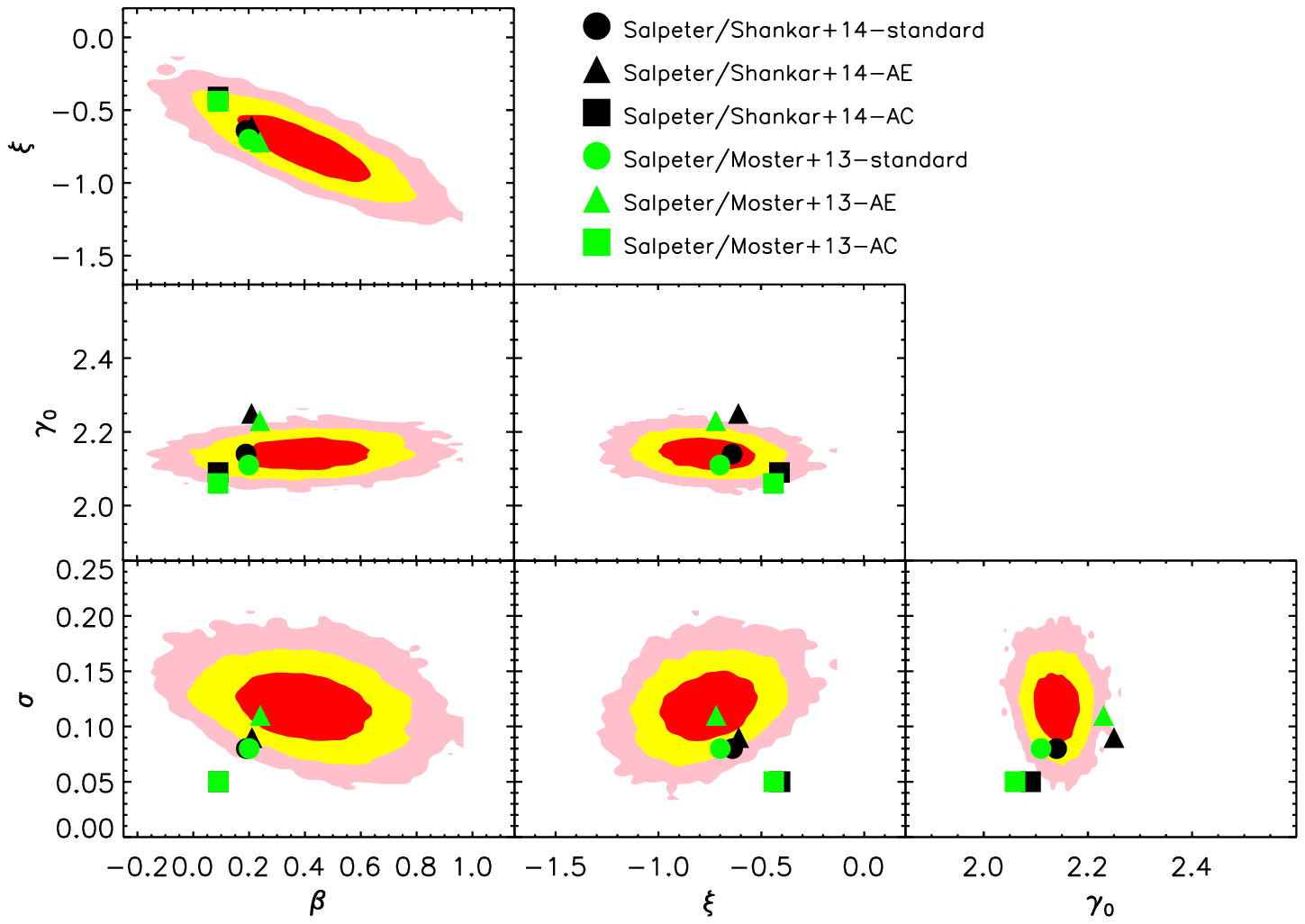}
    \caption{Same format as \figu\ref{fig|GammaPosteriorProbIMF} with the standard model compared with a model characterized by a flatter stellar mass-halo mass relation \citep[][]{Moster13}.
    The latter model tends to be disfavored with respect to a model characterized by a steeper stellar mass-halo mass relation (\figu\ref{fig|MstarMhalo}).
    \label{fig|App:GammaPosteriorProbMoster}}}
\end{figure*}

We here present a series of measured probability distribution functions as extracted from different semi-empirical models and compared with
the data by \citet{Sonne13}. Each Figure has the same format as \figu\ref{fig|GammaPosteriorProbIMF},
with the reference Salpeter IMF model, with stellar \devac\ and dark matter uncontracted NFW profiles,
compared with a model with a variable IMF (\figu\ref{fig|App:GammaPosteriorProbIMFvar}), a Hernquist stellar profile (\figu\ref{fig|App:GammaPosteriorProbHern}),
a core in the dark matter profile as in \citet{Newman13b} of size $r_{\rm core}=14$ kpc
(\figu\ref{fig|App:GammaPosteriorProbCore}), a combination of Hernquist plus cored ($r_{\rm core}=14$ kpc) dark matter profiles (\figu\ref{fig|App:GammaPosteriorProbCoreHern}),
or assuming a reference model characterized by the \citet{Moster13} stellar mass-halo mass relation (\figu\ref{fig|App:GammaPosteriorProbMoster}).
Most (but not all) of these systematic variations to the reference model tend to be disfavored by the data at different significance.

\bibliographystyle{yahapj}
\bibliography{../../../RefMajor_Rossella}

\label{lastpage}
\end{document}